\documentclass[12pt]{article}
\pdfoutput=1
\usepackage{graphicx,psfrag,epsf,color}
\usepackage[dvipsnames]{xcolor}
\usepackage{amsmath,amssymb,amsfonts, booktabs}
\usepackage{appendix}
\usepackage{multirow}
\usepackage{geometry}
\usepackage{float}
\usepackage{mathrsfs}  
\usepackage{authblk}
\usepackage{array}
\usepackage{cite}
\usepackage{slashed}
\usepackage{graphicx}
\graphicspath{ {Img/} }
\usepackage{rotating}
\usepackage{soul}
\usepackage{slashed, cancel}
\usepackage{bbold}
\usepackage[caption=false]{subfig}
\usepackage{hyperref}
\usepackage{comment}
\newcounter{MBQ}

\usepackage{tabularx}
\newcolumntype{C}{>{\centering\arraybackslash}X}

\newcommand{\as}{\alpha_s}
\newcommand{\LamQCD}{\Lambda_{\rm QCD}}
\newcommand{\lb}{\lambda}
\newcommand{\nub}{\bar{\nu}_\ell}

\numberwithin{equation}{section}

\begin{document}
\allowdisplaybreaks

\begin{titlepage}

\begin{flushright}
{\small
January 15, 2025 \\
%arXiv:20mm.nnnnn [hep-ph]
}
\end{flushright}

\vskip1cm
\begin{center}
{\Large \bf\boldmath Kinematic Moments of $\bar{B}\to X_c \ell \bar{\nu}_\ell$\\ to Order $\mathcal{O}(\LamQCD^5/m_b^5)$}
\end{center}

\vspace{0.5cm}
\begin{center}
{\sc Gael Finauri} \\[6mm]
{\it Dipartimento di Fisica, Università di Torino \& INFN, Sezione di Torino,\\
Via Pietro Giuria 1, I-10125 Turin, Italy}

\end{center}

\vspace{0.6cm}
\begin{abstract}
\vskip0.2cm\noindent
We investigate the heavy-quark expansion (HQE) for inclusive semileptonic $\bar{B}$ decays, at tree level, by computing the fully differential decay rate up to order $\mathcal{O}(\Lambda_{\rm QCD}^5/m_b^5)$.
We provide analytic results for the first three moments in the lepton energy, hadronic invariant mass and leptonic invariant mass ($q^2$) with a lower cut on the lepton energy in the $\bar{B}$ rest frame, as well as for the first three $q^2$ moments with a lower cut on $q^2$.
By means of the lowest-lying state saturation ansatz we study the numerical behaviour of the HQE providing insights into the theoretical error budget associated with power corrections.
Available CLEO data on the second central $q^2$ moment with a lower cut on the lepton energy present puzzling discrepancies with the theoretical expectations.
This observation makes new measurements of these observables desirable, which could improve the precision of global fit analyses for the inclusive determination of $|V_{cb}|$.
\end{abstract}
\end{titlepage}

%\email{gael.finauri@unito.it}

%%%%%%%%%%%%%%%%%%%%%%%%%%%%%%%%%%%%%%%%%%%%%%%%%%%%%%%%%%%%%%%%%%%
\tableofcontents
\newpage

\section{Introduction}
For over 20 years, inclusive semileptonic $\bar{B}$ decays have represented one of the main paths in the determination of the Cabibbo-Kobayashi-Maskawa (CKM) matrix element $|V_{cb}|$~\cite{ Battaglia:2002tm,Bauer:2002sh,CLEO:2002woe}.
The fully differential decay rate for $\bar{B}\to X_c \ell \nub$ is computed in the Standard Model through a local Operator Product Expansion (OPE)~\cite{Blok:1993va,Manohar:1993qn,Bigi:1993fe,Gremm:1996df} leveraging the scale hierarchy $m_b \gg \LamQCD$.
The theoretical prediction takes then the form of a double expansion: in $\as$, regarding perturbative QCD corrections, and in $\lb \equiv\LamQCD/m_b$ going under the name of heavy-quark expansion (HQE).
The leading order of the HQE corresponds to the partonic decay $b \to c \ell \nub$, and it is therefore independent on hadronic inputs.
Going beyond the partonic limit, at each order in the HQE new dimensionful non-perturbative parameters arise, encoding the hadronic physics of the inclusive decay.
The goal of global fit analyses~\cite{Bauer:2004ve,Buchmuller:2005zv,Gambino:2013rza,Alberti:2014yda,Bordone:2021oof} is to extract the values of the non-perturbative parameters, crucial for precisely determining $|V_{cb}|$ and all theoretical predictions based on the HQE~\cite{Lenz:2022rbq}, from data.

The result of the OPE for the triple differential rate has to be understood in the sense of a distribution, making the comparison with experimental measurements possible only at the level of integrated quantities.
In the last two decades the first few moments in the lepton energy $E_\ell$ and hadronic invariant mass $m_X^2$, measured by CLEO, DELPHI, CDF, BaBar and Belle~\cite{BaBar:2004bij,CLEO:2004bqt,CDF:2005xlh,DELPHI:2005mot,Belle:2006jtu,Belle:2006kgy,BaBar:2009zpz} have driven the phenomenological studies.
These observables are measured with a lower cut on the lepton energy in the $\bar{B}$ rest frame.
The latest experimental additions were the dilepton invariant mass ($q^2$) moments, as functions of a lower cut $q^2_{\rm cut}$, measured by Belle~\cite{Belle:2021idw} and Belle II~\cite{Belle-II:2022evt} only a few years ago.
The advantage of setting a lower cut in $q^2$, instead of the lepton energy, is that reparametrization invariance (RPI) of the heavy-quark effective theory (HQET) is preserved.
These observables were hence suggested for inclusive fits due to their dependence on a lower set of independent HQE parameters~\cite{Fael:2018vsp}.
The downside is that moments in the lepton energy and the hadronic invariant mass are intrinsically not RPI, and hence cannot be used in fits to RPI quantities~\cite{Bernlochner:2022ucr}.
Combined with increasing efforts in improving the theoretical predictions~\cite{Fael:2020njb,Fael:2020tow,Mannel:2021zzr}, the available measurements allowed for an inclusive determination of $|V_{cb}|$ with a $1.1\%$ relative uncertainty~\cite{Finauri:2023kte}.

From the side of the Standard Model prediction, one of the key aspects is the estimate of theoretical uncertainties.
In order for the prediction to be reliable, they should provide the correct order of magnitude of missing higher order corrections, in both the $\as$ and $\LamQCD/m_b$ expansions.
In this paper we focus on the HQE, at tree level, in order to provide an insight into power suppressed non-perturbative contributions which are difficult to include in global fits due to the proliferation of unknown parameters.
We compute the triple differential distribution for $\bar{B}\to X_c \ell \nub$, and the first three moments in $E_\ell$, $m_X^2$ and $q^2$, up to order $\mathcal{O}(\lb^5)$ in the HQE, checking and updating existing results in the literature~\cite{Mannel:2010wj,Mannel:2023yqf,Fael:2024fkt,Milutin:2024nbd}.
The results of~\cite{Mannel:2023yqf} were recently implemented in the open-source framework \texttt{Kolya}~\cite{Milutin:2024nbd}.
Through the ``lowest-lying state saturation ansatz'' (LLSA)~\cite{Mannel:2010wj,Heinonen:2014dxa} we estimate the magnitude of the non-perturbative parameters appearing at orders $\mathcal{O}(\lb^4)$ and $\mathcal{O}(\lb^5)$, allowing us to study the impact of higher power corrections on the moments of the kinematic distribution.
This study is meant to test the convergence of the HQE and mainly to gain insight in the determination of theoretical uncertainties associated with higher power corrections.

We also investigate the quality of an expansion in $m_c^2/m_b^2 \ll 1$ for all the terms in the OPE, as contributions of the form $\mathcal{O}(\LamQCD^5/(m_c^2 m_b^3))$
appear starting from $\mathcal{O}(\lb^5)$~\cite{Mannel:2010wj,Mannel:2023yqf}, potentially spoiling the power counting.

Furthermore we present for the first time analytic results, up to $\mathcal{O}(\lb^5)$, for the $q^2$ moments computed with a lower cut on the lepton energy.
The first two $q^2$ moments, measured a long time ago by the CLEO collaboration with two different lepton energy cuts, were never employed in global analysis.
While the first moment is in agreement with the theoretical expectation based on the fit~\cite{Finauri:2023kte}, the second central moment presents a rather puzzling tension.
We therefore advocate for new experimental measurements of such observables which could be included in global fits as new interesting constraints.

The paper is organized as follows: in Section~\ref{sec:theory} we review the theoretical framework and main steps of the calculation that lead to the result for the triple differential distribution up to $\mathcal{O}(\lb^5)$.
In Section~\ref{sec:moms} we proceed to compute the $E_\ell$, $m_X^2$ and $q^2$ moments relevant for global fits and comparisons with experimental data.
Our numerical analysis is presented in Section~\ref{sec:num} and we conclude in Section~\ref{sec:summ}.
An appendix contains technical details.

\section{Theoretical Framework}
\label{sec:theory}
The goal of this section is to review the standard derivation of the fully differential decay rate for the inclusive semileptonic decay $\bar{B} \to X_c \ell \bar{\nu}_\ell$.
We start from the effective Hamiltonian, neglecting QED corrections, for the $b \to c \ell \bar{\nu}_\ell$ transition
\begin{equation}
\mathcal{H}_{\rm eff} = \frac{G_F}{\sqrt{2}} V_{cb} [\bar{c} \gamma^\mu (1-\gamma^5) b][\bar{\ell} \gamma_\mu (1-\gamma^5)\nu_\ell]\,,
\end{equation}
where $G_F$ is the Fermi constant and $V_{cb}$ the Cabibbo-Kobayashi-Maskawa (CKM) matrix element.
We define 
\begin{equation}
J_h^\mu(x) = \bar{c}(x) \gamma^\mu P_L b(x)\,, \qquad J_\ell^\mu(x) = \bar{\ell}(x) \gamma^\mu P_L\nu_\ell(x)\,,
\end{equation}
with $P_{L/R} = (1\mp \gamma^5)/2$.
The differential decay rate for the semileptonic inclusive decay is
\begin{align}
d\Gamma(\bar{B}\to X_c \ell \bar{\nu}_\ell) =& \frac{1}{2m_B}\sum_{\text{lept. spins}} \frac{d^3 p_\ell}{(2\pi)^3 2E_\ell}\frac{d^3 p_{\nub}}{(2\pi)^3 2E_{\nub}}\sum_{X_c}(2\pi)^4 \delta^{4}(p_B-p_\ell -p_{\nub}-p_X)\nonumber\\
&\times |\mathcal{A}(\bar{B}\to X_c \ell \bar{\nu}_\ell)|^2\,,
\end{align}
where the sum over the inclusive states $X_c$ encodes the phase space integral.
The amplitude is given by
\begin{equation}
\mathcal{A}(\bar{B} \to X_c \ell \bar{\nu}_\ell) = \langle X_c(p_X) \ell(p_\ell) \nub(p_{\nub})| \mathcal{H}_{\rm eff}|\bar{B}(p_B)\rangle = \frac{4G_F}{\sqrt{2}} V_{cb} \langle X_c| J_h^\mu(0)|\bar{B}\rangle \langle \ell \bar{\nu}_\ell| {J_\ell}_\mu(0)|0\rangle\,,
\end{equation}
where the leptonic and hadronic part factorize, if we neglect QED effects\footnote{QED corrections to inclusive $b \to c\ell \nub$ have been computed in Ref.~\cite{Bigi:2023cbv}}.
We define the leptonic and hadronic tensors as
\begin{align}
L^{\mu\nu} &= \sum_{\text{lept. spins}} \langle 0| {J_\ell^\mu}^\dagger(0) | \ell\nub\rangle \langle \ell \nub |J^\nu_\ell(0) |0\rangle\,,\nonumber\\
W^{\mu\nu} &= \frac{1}{2m_B}\sum_{X_c} (2\pi)^3 \delta^{4}(p_B-q-p_X) \langle \bar{B} | {J_h^\mu}^\dagger(0) |X_c\rangle \langle X_c| J_h^\nu(0)|\bar{B}\rangle
\end{align}
where we defined $q=p_\ell + p_{\nub}$.
In this way we write the differential decay rate as
\begin{equation}
\label{eq:dGammagen}
d\Gamma(\bar{B}\to X_c \ell \nub) =16\pi G_F^2 |V_{cb}|^2 \frac{d^3 p_\ell}{(2\pi)^3 2E_\ell}\frac{d^3 p_{\nub}}{(2\pi)^3 2E_{\nub}}L^{\mu\nu}W_{\mu\nu}\,.
\end{equation}
The leptonic tensor can be computed right away~\cite{Manohar:2000dt}
\begin{equation}
L^{\mu\nu} = 2\bigl( p_\ell^\mu p_{\nub}^\nu + p_\ell^\nu p_{\nub}^\mu -g^{\mu\nu} p_\ell \cdot p_{\nub} + i \epsilon^{\mu\nu\alpha \beta} {p_\ell}_\alpha {p_{\nub}}_\beta\bigr)\,,
\end{equation}
with the convention $\epsilon^{0123}=-\epsilon_{0123} = +1$.
On the other hand the hadronic tensor $W^{\mu\nu}$ can be decomposed into five scalar functions as
\begin{equation}
\label{eq:Wtensdec}
m_b W^{\mu\nu} = -W_1 g^{\mu\nu} +W_2 v^\mu v^\nu +i W_3 \epsilon^{\mu\nu \rho \sigma} v_\rho \hat{q}_\sigma + W_4 \hat{q}^\mu \hat{q}^\nu + W_5(\hat{q}^\mu v^\nu + \hat{q}^\nu v^\mu)\,,
\end{equation}
with $\hat{q}^\mu = q^\mu/m_b$ and $v^\mu = p_B^\mu/m_B$, so that the structure functions $W_i$ are dimensionless.
With this the product between leptonic and hadronic tensor reads
\begin{equation}
\label{eq:LtimesW}
\frac{1}{m_b}L^{\mu\nu}W_{\mu\nu} = 2\hat{q}^2 W_1 + (4\hat{E}_\ell \hat{E}_{\nub}-\hat{q}^2)W_2 + 2\hat{q}^2(\hat{E}_\ell - \hat{E}_{\nub})W_3\,,
\end{equation}
where we normalized the charged lepton and neutrino energies by $m_b$, and we neglected the lepton mass. In this limit, the structure functions $W_4$ and $W_5$ do not contribute to the decay rate.
Inserting~\eqref{eq:LtimesW} into~\eqref{eq:dGammagen} we find
\begin{align}
\frac{d^3\Gamma}{d\hat{E}_\ell d\hat{E}_{\nub} d\cos\theta} =&\; 96\Gamma_0 \theta(\hat{E}_\ell)\theta(\hat{E}_{\nub}) \theta(1-\cos\theta)\theta(1+\cos\theta) \hat{E}_\ell \hat{E}_{\nub}\nonumber\\
&\times\biggl[2\hat{q}^2 W_1 + (4\hat{E}_\ell \hat{E}_{\nub}-\hat{q}^2)W_2 + 2\hat{q}^2(\hat{E}_\ell - \hat{E}_{\nub})W_3 \biggr]\,,
\end{align}
where we have to consider $\hat{q}^2 = 2\hat{E}_\ell \hat{E}_{\nub}(1-\cos\theta)$ and we defined
\begin{equation}
\Gamma_0 = |V_{cb}|^2\frac{G_F^2 m_b^5}{192\pi^3}\,.
\end{equation}
Changing variable from $\cos\theta$ to $\hat{q}^2$ we get
\begin{align}
\frac{d^3\Gamma}{d\hat{E}_\ell d\hat{E}_{\nub} d\hat{q}^2} =&\; 48\Gamma_0 \theta(\hat{E}_\ell) \theta(\hat{q}^2)\theta(4\hat{E}_\ell \hat{E}_{\nub}-\hat{q}^2) \nonumber\\
&\times\biggl[2\hat{q}^2 W_1 + (4\hat{E}_\ell \hat{E}_{\nub}-\hat{q}^2)W_2 + 2\hat{q}^2(\hat{E}_\ell - \hat{E}_{\nub})W_3 \biggr]\,,
\end{align}
where we were able to remove $\theta(\hat{E}_{\nub})$ as it is implied by the other three Heaviside functions.
Finally defining $\hat{u} = ((m_b v-q)^2-m_c^2)/m_b^2$, the mass ratio $\rho =m_c^2/m_b^2$, and changing variable through $\hat{E}_{\nub} = (1-\rho+\hat{q}^2-2\hat{E}_\ell -\hat{u})/2$ we get
\begin{align}
\label{eq:d3Gamma}
\frac{d^3 \Gamma}{d\hat{E}_\ell \,d\hat{u}\, d\hat{q}^2} =&\; 48\Gamma_0 \theta(\hat{E}_\ell) \theta(\hat{q}^2) \theta\bigl(-4\hat{E}_\ell^2 + 4\hat{q}_0 \hat{E}_\ell-\hat{q}^2\bigr)\nonumber\\
&\times \biggl[\hat{q}^2 W_1 - \Bigl(2\hat{E}_\ell^2 - 2\hat{E}_\ell \hat{q}_0 + \frac{\hat{q}^2}{2} \Bigr) W_2 + \hat{q}^2 (2\hat{E}_\ell - \hat{q}_0)W_3 \biggr]\,,
\end{align}
where we used the useful combination $\hat{q}_0 = (1-\rho +\hat{q}^2-\hat{u})/2$. 
The advantage of this final expression is that the structure functions $W_i$ are independent on the lepton energy $\hat{E}_\ell$.
In order to compute moments of the spectrum we now need to calculate explicitly the structure functions, which we will do at leading order in the perturbative expansion and to order $\mathcal{O}(\lb^5)$ in the so called heavy-quark expansion (HQE).

Before moving on to the next section, we first recall some definitions of the heavy-quark effective theory (HQET) which will be used in the calculation of $W^{\mu\nu}$.
We define the projectors
\begin{equation}
P_\pm = \frac{1\pm \slashed{v}}{2}\,,
\end{equation}
with $v$ the heavy quark velocity four-vector ($v^2=1$), such that $P_+ P_- = P_- P_+ = 0$, $P_\pm^2 = P_\pm$ and $P_+ + P_- = 1$.
We will work with the rephased full QCD field
\begin{equation}
\label{eq:bvfield}
b_v(x) \equiv e^{i m_b v \cdot x} b(x)\,,
\end{equation}
such that a covariant derivative acting on it can be considered to be of order $\mathcal{O}(\LamQCD)$.
It is straightforward from~\eqref{eq:bvfield} to derive the following equation of motion
\begin{equation}
\label{eq:bveom1}
P_- b_v(x) = \frac{i \slashed{D}}{2m_b}b_v(x)\,,
\end{equation}
which will be useful in manipulating operators to determine the correct scaling in the $\lb$ expansion.
From~\eqref{eq:bveom1}, applying $i\slashed{D}$ to both sides of the equation, we can derive
\begin{equation}
\label{eq:bveom2}
(i v\cdot D) b_v = -\frac{1}{2m_b}(i\slashed{D})(i\slashed{D})b_v\,,
\end{equation}
which also relates operators of different dimensions.
The relation~\eqref{eq:bveom2} shows that, when acting on the $b_v$ field, a covariant derivative in the direction of the heavy quark velocity is power suppressed.
It is hence customary to split the parallel and perpendicular components
\begin{equation}
    iD^\mu = iD^\mu_\perp + v^\mu (i v\cdot D)\,,
\end{equation}
which can be done in general for any vector by introducing
\begin{equation}
g^{\mu\nu}_\perp = g^{\mu\nu}-v^\mu v^\nu\,, \qquad \epsilon^{\mu \nu \rho}_\perp = \epsilon^{\mu \nu \rho \sigma}v_\sigma\,,
\end{equation}
such that $g^{\mu\nu}_\perp v_\mu = g^{\mu\nu}_\perp v_\nu = 0$, $g_\perp^{\mu\nu}{g_{\perp}}_{\mu\nu} = 3$ and $\epsilon_\perp^{\mu\nu\rho}v_\rho=0$.
A particularly useful relation for manipulating spin-dependent operators is
\begin{equation}
\gamma^\mu_\perp \gamma^5 = -\frac{\epsilon^{\mu \alpha \beta}_\perp}{2} \sigma_{\alpha\beta} \slashed{v}\,,
\end{equation}
where $\gamma^\mu_\perp = g^{\mu\nu}_\perp \gamma_\nu$ and $\sigma_{\alpha\beta} = \frac{i}{2}[\gamma_\alpha,\gamma_\beta]$. In the following we will also use $\sigma_\perp^{\alpha\beta} = g_\perp^{\alpha\mu}g_\perp^{\beta\nu}\sigma_{\mu\nu}$.

\subsection{Operator Product Expansion}
\label{sec:OPE}
The hadronic tensor can be computed through an OPE starting from the following forward matrix element~\cite{Manohar:1993qn}
\begin{equation}
T^{\mu\nu}(q) \equiv \frac{i}{2m_B}\int d^4 x e^{-iq \cdot x} \langle \bar{B} | \text{T}\{{J_h^\mu}^\dagger(x) J_h^\nu(0) \} |\bar{B}\rangle\,,
\end{equation}
which is related to $W^{\mu\nu}$ through the optical theorem~\cite{Manohar:1993qn, Manohar:2000dt}
\begin{equation}
\label{eq:WImT}
W^{\mu\nu} = \frac{1}{\pi} \text{Im}[T^{\mu\nu}]\,.
\end{equation}
In analogy with~\eqref{eq:Wtensdec} we write down the tensor decomposition
\begin{equation}
\label{eq:Ttensdec}
m_b T^{\mu\nu} = -T_1 g^{\mu\nu} +T_2 v^\mu v^\nu +i T_3 \epsilon^{\mu\nu \rho \sigma} v_\rho \hat{q}_\sigma + T_4 \hat{q}^\mu \hat{q}^\nu + T_5(\hat{q}^\mu v^\nu + \hat{q}^\nu v^\mu)\,,
\end{equation}
where $T_4$ and $T_5$ only contribute for non-vanishing lepton mass.
To compute the imaginary part according to~\eqref{eq:WImT}, we will use the relation
\begin{equation}
\frac{1}{\pi}\text{Im}\biggl[\frac{1}{(\omega + i\eta)^{n+1}}\biggr] = \frac{(-1)^{n+1}}{n!}\delta^{(n)}(\omega)\,,
\end{equation}
where $\delta^{(n)}(\omega)$ is the $n$-th derivative of $\delta(\omega)$ with respect to $\omega$.
The tensor $T^{\mu\nu}$ is computed through an OPE using, at tree level, the background field method for the charm quark propagator~\cite{Novikov:1984ecy, Blok:1993va, Mannel:2010wj}
\begin{align}
T^{\mu\nu}(q) =& \frac{i}{2m_B} \int d^4x e^{i (m_b v-q) \cdot x}\langle \bar{B} | \text{T}\{\bar{b}_v(x) \gamma^\mu P_L c(x) \bar{c}(0) \gamma^\nu P_L b_v(0)\}|\bar{B}\rangle\nonumber\\
=& \frac{i}{2m_B} \int d^4x e^{i (m_b v-q) \cdot x}\langle \bar{B} | \bar{b}_v(x) \gamma^\mu P_L S_{\rm BGF}(x,0) \gamma^\nu P_L b_v(0)|\bar{B}\rangle\,,
\end{align}
where the second line has been expanded for small $x$, with
\begin{equation}
S_{\rm BGF}(x,0) = i \int \frac{d^4 p}{(2\pi)^4}e^{-i p\cdot x}\frac{\slashed{p}+m_c}{p^2-m_c^2+i\eta}\sum_{n=0}^\infty (-1)^n\Bigl[i \slashed{D} \frac{\slashed{p}+m_c}{p^2-m_c^2+i\eta} \Bigr]^n\,.
\end{equation}
Performing the integration over $x$, and defining $p^\mu = m_b v^\mu-q^\mu$, we get
\begin{equation}
T^{\mu\nu}(q) = -\frac{1}{2m_B}\sum_{n=0}^\infty \frac{(-1)^n}{(p^2-m_c^2+i\eta)^{n+1}} \langle \bar{B} |\bar{b}_v(0) \gamma^\mu P_L (\slashed{p}+m_c)[i \slashed{D}(\slashed{p}+m_c)]^n \gamma^\nu P_L  b_v(0) |\bar{B} \rangle\,,
\end{equation}
and from now on we will omit the argument of the fields when evaluated at position $x=0$.
We define the forward matrix element of a general operator of dimension $n+3$ through the following projector~\cite{Mannel:2010wj}
\begin{equation}
\label{eq:Mndef}
M^{(n)\mu_1 ... \mu_n}_{ij}(v) =\frac{1}{2m_B} \langle \bar{B}| \bar{b}_{v,j} (iD^{\mu_1})...(i D^{\mu_n}) b_{v,i}|\bar{B}\rangle\,,
\end{equation}
where $i,j$ are spinor indices and the external states have four-momentum $p_B^\mu=m_B v^\mu$, so that
\begin{equation}
\label{eq:Tstep}
T^{\mu\nu}(q) = \sum_{n=0}^\infty \frac{(-1)^{n+1}}{(p^2-m_c^2+i\eta)^{n+1}} \text{tr}\biggl\{\gamma^\nu P_L M^{(n)}_{\mu_1...\mu_n}P_R \gamma^\mu (\slashed{p}+m_c)\Bigl[\prod_{k=1}^n \gamma^{\mu_k}(\slashed{p}+m_c)\Bigr] \biggr\}\,.
\end{equation}
We can now expand~\eqref{eq:Mndef} in the usual Dirac matrices basis
\begin{equation}
\label{eq:Mngeneral}
[M^{(n)}_{\mu_1...\mu_n}(v)]_{ij} =\frac{1}{4}\biggl[ S^{(n)}_{\mu_1 ...\mu_n} \mathbb{1}_{ij} + P^{(n)}_{\mu_1...\mu_n}  [\gamma^5]_{ij} + V^{(n)\alpha}_{\mu_1...\mu_n} [\gamma_\alpha]_{ij} +A^{(n)\alpha}_{\mu_1...\mu_n}  [i \gamma_\alpha \gamma^5]_{ij} + \frac{T^{(n)\alpha \beta}_{\mu_1...\mu_n}}{\sqrt{2}} [\sigma_{\alpha\beta}]_{ij}\biggr]\,,
\end{equation}
where the normalization is such that the scalar functions are obtained with
\begin{align}
S^{(n)}_{\mu_1...\mu_n} &= \text{tr}\{M^{(n)}_{\mu_1...\mu_n}\}\,,\qquad P^{(n)}_{\mu_1...\mu_n} = \text{tr}\{M^{(n)}_{\mu_1...\mu_n}\gamma^5\}\,,\qquad V^{(n)\alpha}_{\mu_1...\mu_n} = \text{tr}\{M^{(n)}_{\mu_1...\mu_n}\gamma^\alpha\}\,, \nonumber\\
A^{(n)\alpha}_{\mu_1...\mu_n} &= \text{tr}\{M^{(n)}_{\mu_1...\mu_n}i\gamma^\alpha \gamma^5\}\,, \qquad T^{(n)\alpha \beta}_{\mu_1...\mu_n} = \frac{1}{\sqrt{2}}\text{tr}\{M^{(n)}_{\mu_1...\mu_n}\sigma^{\alpha \beta}\}\,.
\end{align}
Using the expansion~\eqref{eq:Mngeneral} in~\eqref{eq:Tstep} we arrive at a final expression
\begin{equation}
\label{eq:Tmunufinal}
m_b T^{\mu\nu}(q) = \sum_{n=0}^\infty \biggl(V^{(n)\alpha}_{\mu_1...\mu_n} + i A^{(n)\alpha}_{\mu_1...\mu_n} \biggr) {G^{(n)}_\alpha}^{\mu\nu \mu_1 ... \mu_n}\,,
\end{equation}
with
\begin{equation}
{G^{(n)}_\alpha}^{\mu\nu\mu_1...\mu_n} = \frac{(-1)^{n+1}}{4m_b^{2n+1}(\hat{u}+i\eta)^{n+1}}\text{tr}\biggl\{P_R \gamma^\nu \gamma_\alpha \gamma^\mu (\slashed{p}+m_c)\Bigl[\prod_{k=1}^n \gamma^{\mu_k}(\slashed{p}+m_c)\Bigr] \biggr\}\,,
\end{equation}
which can be easily computed to each order in $n$ with available \texttt{Mathematica} codes~\cite{Shtabovenko:2020gxv}.
This reduces the calculation of the triple differential spectrum~\eqref{eq:d3Gamma} up to order $\mathcal{O}(\lb^5)$ to the determination of $M^{(n)}_{\mu_1...\mu_n}$ up to $n=5$ .

\subsection{Determination of $M^{(n)}_{\mu_1...\mu_n}(v)$ up to $n=5$}
The calculation of the projector $M^{(n)}_{\mu_1...\mu_n}(v)$ is algorithmic~\cite{Mannel:2010wj}, following from the equations of motion for the $b_v$ field~\eqref{eq:bveom1}-\eqref{eq:bveom2}.
As a first simplification, parity invariance of QCD implies the following constraint (no summation over indices)
\begin{equation}
\label{eq:Pconst}
M^{(n)\mu_1...\mu_n}(v) = \tilde{I}^{\mu_1} ... \tilde{I}^{\mu_n} [\gamma^0 M^{(n)\mu_1...\mu_n}(\tilde{v}) \gamma^0]
\end{equation}
where $\tilde{I}^\mu = (1,-1,-1,-1)$ and $\tilde{v}^\mu = (v^0, -v^1, -v^2, -v^3)$.
By using (no sum)
\begin{equation}
\tilde{I}^\mu \tilde{v}^\mu = v^\mu\,, \qquad \tilde{I}^\mu \tilde{I}^\nu g^{\mu\nu} = g^{\mu\nu}\,, \qquad \tilde{I}^\mu \tilde{I}^\nu \tilde{I}^\rho \tilde{I}^\sigma \epsilon^{\mu\nu\rho\sigma} = -\epsilon^{\mu\nu\rho\sigma}\,,
\end{equation}
and equation~\eqref{eq:Pconst}, we arrive at the conclusion that $S^{(n)}$, $V^{(n)}$ and $T^{(n)}$ cannot contain a Levi-Civita tensor (we remind that a combination of two Levi-Civita tensors can always be rewritten as combinations of the metric tensor), while $P^{(n)}$ and $A^{(n)}$ must contain the Levi-Civita tensor to satisfy the parity constraint.

Furthermore we show that some of the functions multiplying the independent Dirac structures in~\eqref{eq:Mngeneral} can be rewritten in terms of traces of $M^{(n+1)}$.
By using the identities $\slashed{v} = 1-2P_-$, $\gamma^5 = P_- \gamma^5 + \gamma^5 P_-$ and $\gamma^\alpha_\perp = P_- \gamma^\alpha_\perp + \gamma^\alpha_\perp P_-$ we obtain
\begin{align}
\label{eq:recurrence}
V^{(n)\alpha}_{\mu_1...\mu_n} &= v^\alpha S^{(n)}_{\mu_1...\mu_n} - \frac{v^\alpha}{m_b} \text{tr}\Bigl\{ M^{(n+1)}_{\mu_1...\mu_n \rho} \gamma^\rho \Bigr\} + \frac{1}{2m_b}\text{tr}\Bigl\{ M^{(n+1)}_{\mu_1...\mu_n \rho} \gamma^\alpha_\perp \gamma^\rho \Bigr\}+ \frac{1}{2m_b}\text{tr}\Bigl\{ M^{(n+1)}_{\rho\mu_1...\mu_n} \gamma^\rho \gamma^\alpha_\perp \Bigr\}\,,\nonumber\\
P^{(n)}_{\mu_1...\mu_n} &= \frac{1}{2m_b}\text{tr}\Bigl\{(M^{(n+1)}_{\rho\mu_1...\mu_n} -M^{(n+1)}_{\mu_1...\mu_n\rho})\gamma^\rho \gamma^5 \Bigr\}\,,\nonumber\\
A^{(n)\alpha}_{\mu_1...\mu_n} &= \frac{\epsilon^{\alpha \mu \nu}_\perp}{2} t^{(n)}_{\mu\nu\mu_1...\mu_n} + \frac{i}{m_b}\text{tr}\Bigl\{M^{(n+1)}_{\mu_1...\mu_n \rho} \slashed{v}\gamma_\perp^\alpha \gamma^\rho \gamma^5 \Bigr\} -\frac{iv^\alpha}{2m_b}\text{tr}\Bigl\{(M^{(n+1)}_{\mu_1...\mu_n \rho} + M^{(n+1)}_{\rho \mu_1...\mu_n})\gamma^\rho \gamma^5 \Bigr\}\,,\nonumber\\
T^{(n)\alpha\beta}_{\mu_1...\mu_n} &=\frac{i}{\sqrt{2}}\biggl[t^{(n)\alpha\beta}_{\mu_1...\mu_n} -\frac{v^\alpha}{2m_b}\text{tr}\Bigl\{M^{(n+1)}_{\rho \mu_1 ...\mu_n} \gamma^\rho \gamma^\beta_\perp - M^{(n+1)}_{\mu_1...\mu_n \rho} \gamma^\beta_\perp \gamma^\rho \Bigr\}\nonumber\\
&\qquad\qquad\qquad +\frac{v^\beta}{2m_b}\text{tr}\Bigl\{M^{(n+1)}_{\rho \mu_1 ...\mu_n} \gamma^\rho \gamma^\alpha_\perp - M^{(n+1)}_{\mu_1...\mu_n \rho} \gamma^\alpha_\perp \gamma^\rho \Bigr\}  \biggr]\,,
\end{align}
therefore we only need to tensor decompose, taking into account only terms without the Levi-Civita tensor (from the parity constraint), the two independent functions
\begin{align}
S^{(n)}_{\mu_1...\mu_n} &= \frac{1}{2m_B}\langle \bar{B} |\bar{b}_v iD_{\mu_1} ... iD_{\mu_n} b_v|\bar{B}\rangle\,,\nonumber\\
t^{(n)\alpha\beta}_{\mu_1...\mu_n} &= \frac{1}{2m_B}\langle \bar{B} |\bar{b}_v iD_{\mu_1} ... iD_{\mu_n}(-i \sigma^{\alpha\beta}_\perp) b_v|\bar{B}\rangle\,,
\end{align}
which will be parametrized in terms of the HQE parameters defined in Appendix~\ref{sec:appHQE}~\cite{Mannel:2010wj}.
These relations allow to virtually compute power corrections to any order in $1/m_b$ (at order $\mathcal{O}(\as^0)$) once a basis of independent operators is identified at each order.
In this work we choose to stop at $1/m_b^5$ as higher power corrections are far from being useful to phenomenological applications.
This means that in our calculation $M^{(6)}_{\mu_1 ...\mu_6} = 0$, giving for $M^{(5)}_{\mu_1...\mu_5}$ the static limit result
\begin{equation}
    M^{(5)}_{\mu_1...\mu_5}(v) = \frac{1}{2}S^{(5)}_{\mu_1...\mu_5} P_+ + \frac{i}{4}\sigma_{\perp\alpha \beta} P_+ t^{(5)\alpha\beta}_{\mu_1...\mu_5} \,.
\end{equation}
Our results for $M^{(n)}_{\mu_1...\mu_n}$ agree with the results provided in the ancillary files from Ref.~\cite{Mannel:2023yqf} up to order $\mathcal{O}(\lb^4)$.
At order $\mathcal{O}(\lb^5)$ we find a difference which is always proportional to the following RPI combination of HQE parameters
\begin{equation}
\label{eq:mb5diff}
    r_9 + r_{10} - r_{11} - r_{12} + 2 r_{13} - r_{15} - r_{16} + r_{18} = -X_8^5 + \frac{1}{2}X_{10}^5\,,
\end{equation}
where $X_i^5$ are the RPI HQE parameters\footnote{Appendix A of~\cite{Mannel:2023yqf} presents conversion formulas between the $X_i^5$ and $r_i$ basis. We checked those conversion formulas and agree with (A.2) of~\cite{Mannel:2023yqf}.} defined in~\cite{Mannel:2023yqf}.

The difference likely comes from the fact that the relation
\begin{equation}
\epsilon^{\rho \alpha \beta}_\perp A^{(4)}_{\rho \mu_1 ...\mu_4} = i \sqrt{2} g_\perp^{\alpha \rho} g_\perp^{\beta \delta} T^{(4)}_{\rho\delta \mu_1 ...\mu_4}\,,
\end{equation}
implied by~\eqref{eq:recurrence}, is not satisfied by the results presented in~\cite{Mannel:2023yqf}.

With the explicit expression for $M^{(n)}_{\mu_1...\mu_n}$ we determine the scalar functions $T_i$ through~\eqref{eq:Tmunufinal} and by taking the imaginary part according to~\eqref{eq:WImT} we find the structure functions $W_i$ entering the triple differential decay rate.
After some simplifications, the dependence of the structure functions on the kinematic variables $\hat{u}$ and $\hat{q}^2$ can be reduced to the form
\begin{equation}
\label{eq:Wstruct}
W_i(\hat{q}^2,\hat{u}) = \sum_{n=0}^\infty \sum_{k=0}^{n+1} w_i^{(n,k)} \hat{q}^{2k} \delta^{(n)}(\hat{u})\,,
\end{equation}
which holds to all orders in $n$. The coefficients $w_i^{(n,k)}$ are functions of the non-perturbative HQE parameters of mass dimension $\geq n$.
We provide our analytic results for $W_{1,...,5}$ in the ancillary files.

\section{Moments of the Semileptonic Spectrum}
\label{sec:moms}
In this section we compute moments of the semileptonic decay rate starting from the triple differential distribution~\eqref{eq:d3Gamma}.
The three types of moments, in $q^2$, lepton energy $E_\ell$ and hadronic invariant mass $m_X^2$, can be obtained simultaneously if computed with a lower cut ${E_\ell}_{\rm cut}$ on the lepton energy.
On the other hand, $q^2$ moments with a lower cut $q^2_{\rm cut}$, preserving RPI, have to be computed separately, which we will do in Section~\ref{sec:q2momcut}.

\subsection{All Moments with Cut on the Lepton Energy}
\label{sec:allmom}
Moments computed with only a lower cut $\hat{E_\ell}_{\rm cut} = {E_\ell}_{\rm cut}/m_b$ can be computed simultaneously and in complete generality.
This is due to the fact that the structure functions at tree level are polynomial in $\hat{q}^2$ and depend on $\hat{u}$ only through delta functions and their derivatives as shown in~\eqref{eq:Wstruct}.
Hence we can write down the triple differential distribution~\eqref{eq:d3Gamma} in the form
\begin{equation}
\frac{d^3\Gamma}{dy\, d\hat{u}\, d\hat{q}^2} = \Gamma_0 \theta(\hat{q}^2) \theta(y) \theta(g(y,\hat{u})-\hat{q}^2) \sum_{n=0}^\infty \sum_{i'=0}^2 \sum_{j'=0}^1 \sum_{k'=0}^{n+2} \gamma^{(n)}_{i'j'k'} y^{i'} \hat{u}^{j'} \hat{q}^{2k'} \delta^{(n)}(\hat{u})\,,
\end{equation}
where $y=2\hat{E}_\ell$ and the coefficients $\gamma^{(n)}_{i' j' k'}$ are determined as combinations of the $w_i^{(n,k')}$ coefficients from the expression for the differential rate~\eqref{eq:d3Gamma}.
We defined the function
\begin{equation}
g(y,\hat{u}) = \frac{y}{1-y}(1-\rho-y-\hat{u})\,.
\end{equation}
The moments with a cut on the lepton energy $y_{\rm cut} = 2\hat{E_\ell}_{\rm cut}$ can be computed as
\begin{align}
M_{ijk}(y_{\rm cut}) =& \frac{1}{2^i\Gamma_0}\int_{y_{\rm cut}}^\infty dy \int d\hat{u}\, d\hat{q}^2 y^i \hat{u}^j \hat{q}^{2k} \frac{d^3\Gamma}{dy\, d\hat{u}\, d\hat{q}^2}\nonumber\\
=& \frac{1}{2^i}\sum_{n=0}^\infty \sum_{i',j',k'} \gamma^{(n)}_{i'j'k'} \int_{y_{\rm cut}}^\infty dy\, y^{i+i'}  \int_{-\infty}^{+\infty} d\hat{u}\, \delta^{(n)}(\hat{u}) \hat{u}^{j+j'}\int_0^{g(y,\hat{u})}d\hat{q}^2  \hat{q}^{2(k+k')}\nonumber\\
=& \frac{1}{2^i}\sum_{n=0}^\infty \sum_{i',j',k'} \gamma^{(n)}_{i'j'k'} \int_{y_{\rm cut}}^\infty dy\, \frac{y^{i+i'}}{k+k'+1}  \int_{-\infty}^{+\infty} d\hat{u}\, \theta(1-\rho-y-\hat{u})\delta^{(n)}(\hat{u}) \hat{u}^{j+j'}  g(y,\hat{u})^{k+k'+1}\,,
\end{align}
where now the integral on $\hat{u}$ can be performed using integration by parts as 
\begin{equation}
\int_{-\infty}^{+\infty} d\hat{u} \,\delta^{(n)}(\hat{u}) f(\hat{u}) = (-1)^n f^{(n)}(\hat{u})\bigl|_{\hat{u}=0}\,.
\end{equation}
So the result can be cast in the following form
\begin{align}
M_{ijk}(y_{\rm cut}) = \frac{1}{2^i} \sum_{a,b,c_k} \int_{y_{\rm cut}}^\infty dy\, \frac{y^{i+a}}{(1-y)^b} g(y,0)^{c_k}\Bigl[ \theta(1-\rho-y) h^{a b c_k}_{jk} +\sum_{n=0}^{\infty} d_{njk}^{abc_k}\delta^{(n)}(y-1+\rho) \Bigr]\,,
\end{align}
where the new coefficients $h^{abc_k}_{jk}$ and $d^{abc_k}_{njk}$ are easily expressed as functions of $\gamma^{(n)}_{i'j'k'}$ and ultimately of $w_i^{(n,k')}$.
Notice that the values of the index $c_k$ depend on the initial value of $k$.
The final result is hence expressed in terms of a single family of master integrals
\begin{equation}
I(a,b,c,y_{\rm cut}) \equiv \int_{y_{\rm cut}}^{1-\rho} dy\, \frac{y^a}{(1-y)^b}(1-\rho-y)^c\,,
\end{equation}
which is in general expressed in closed form through an Appel function.
However for integer values of the parameters, which is our case, the integral can be further decomposed as
\begin{equation}
I(a,b,c,y_{\rm cut}) = \int_{y_{\rm cut}}^{1-\rho}dy\, \frac{y^{a}}{(1-y)^b}\sum_{l=0}^c (-\rho)^l {c \choose l}(1-y)^{c-l} = \sum_{l=0}^c (-\rho)^l {c \choose l} I(a,b-c+l,0,y_{\rm cut})\,,
\end{equation}
which is expressed in terms of simpler hypergeometric functions, resulting after explicit evaluation to combination of logarithms.
Hence the general moments at tree level are given by
\begin{align}
M_{ijk}(y_{\rm cut}) =& \frac{1}{2^i} \sum_{a,b,c_k}\biggl\{ I(i+a+c_k,b+c_k,c_k,y_{\rm cut}) h^{a b c_k}_{jk}\nonumber\\
&+\sum_{n=0}^{\infty} d_{njk}^{abc_k}(-1)^n \frac{d^n}{dy^n} \biggl[\frac{y^{i+a}}{(1-y)^b} g(y,0)^{c_k}\biggr]\biggl|_{y=1-\rho} \biggr\}\,,
\end{align}
with the sum cut at $n=5$ in our case.
We stress that the only knowledge required in order to have a final analytic closed form result for the moments $M_{ijk}$ is the expression of the coefficients $w_i^{(n,k)}$ of the structure functions in terms of the HQE parameters.
In the following sections we define explicitly moments in the lepton energy, hadronic invariant mass and dilepton invariant mass ($q^2$) which are the fundamental observables for the inclusive determination of $|V_{cb}|$~\cite{Gambino:2013rza}.

\subsection{$E_\ell$ Moments}
Moments in the lepton energy are measured with a lower cut ${E_\ell}_{\rm cut}$ and therefore already included in the results from Section~\ref{sec:allmom}.
We have cross-checked the calculation from scratch following a different integration order~\cite{Manohar:2000dt}. 
The results of the linear moments, explicitly defined as
\begin{equation}
\mathcal{L}_n(\hat{E_\ell}_{\rm cut}) \equiv \frac{1}{\Gamma_0} \int_{\hat{E_\ell}_{\rm cut}}^{(1-\rho)/2}d \hat{E}_\ell (\hat{E}_\ell)^n \frac{d\Gamma}{d\hat{E}_\ell} = M_{n00}(2\hat{E_\ell}_{\rm cut})\,,
\end{equation}
for $n=0,1,2,3$ are given in the \texttt{Mathematica} files \texttt{L$n$.m} in the folder \texttt{ResultsElMom/} of the ancillary files.
The results agree up to $\mathcal{O}(\lb^4)$ with the previous determination~\cite{Mannel:2010wj}, while they differ at order $\mathcal{O}(\lb^5)$.
This should not be worrying as a mistake in the derivation of~\cite{Mannel:2010wj} was pointed out in~\cite{Mannel:2023yqf}.

\subsection{$m_X^2$ Moments}
Moments in the hadronic invariant mass are defined as 
\begin{equation}
\mathcal{H}_n(\hat{E_\ell}_{\rm cut}) = \frac{1}{\Gamma_0} \int_{\hat{E_\ell}_{\rm cut}}^{\infty} d\hat{E_\ell} \int d\hat{u}\, d\hat{q}^2 (\hat{m}_X^2)^n \frac{d^3 \Gamma}{d\hat{E}_\ell d\hat{u}\,d\hat{q}^2}\,,
\end{equation}
where the normalized hadronic invariant mass is given by
\begin{equation}
\hat{m}^2_X = \frac{1}{m_b^2}(p_B-q)^2 = \hat{\bar{\Lambda}}^2+(1+\rho)\hat{\bar{\Lambda}} +\rho +(1+\hat{\bar{\Lambda}}) \hat{u} - \hat{\bar{\Lambda}} \hat{q}^2\,,
\end{equation}
with the parameter
\begin{equation}
\hat{\bar{\Lambda}} = \frac{1}{m_b}(m_B - m_b)\,,
\end{equation}
which we count as $\mathcal{O}(1)$ as done in previous analyses~\cite{Gambino:2013rza,Gambino:2011cq}.
The results are obtained as linear combinations of the building blocks $M_{0jk}(2\hat{E_\ell}_{\rm cut})$ calculated in Section~\ref{sec:allmom}.
The explicit expressions for $n=0,1,2,3$ are given in the \texttt{Mathematica} files \texttt{H$n$.m} in the folder \texttt{ResultsHMom/} of the ancillary files.
Again the results agree up to $\mathcal{O}(\lb^4)$ with~\cite{Mannel:2010wj}, with differences of order $\mathcal{O}(\lb^5)$ as expected.

\subsection{$q^2$ Moments}
\label{sec:q2momcut}
Moments in the dilepton invariant mass have been measured with a lower cut on the lepton energy by the CLEO collaboration roughly 20 years ago.
With the results from Section~\ref{sec:allmom} they are easily expressed as
\begin{equation}
    \mathcal{Q}_n(\hat{E_\ell}_{\rm cut}) = M_{00n}(2\hat{E_\ell}_{\rm cut})\,,
\end{equation}
which we present, for $n=0,1,2,3$, in the \texttt{Mathematica} ancillary files \texttt{QElcut$n$.m} in the folder \texttt{Resultsq2ElcutMom/}.
Interestingly these observables where never included in global fits for the extraction of $|V_{cb}|$, and deserve a separate discussion which we postpone to Section~\ref{sec:CLEO}.
The disadvantage of a lower cut in the lepton energy is that it leads to the breaking of RPI. In the case of $E_\ell$ and $m_X^2$ moments this does not pose any problem since these quantities are already non-RPI from the start.
On the other hand $q^2$ moments are potentially RPI observables, if a suitable cut is chosen.
This is the reason why $q^2$ moments with a lower cut on $q^2$ and no cut on the lepton energy were introduced~\cite{Fael:2018vsp}, preserving RPI and depending on a restricted set of HQE parameters~\cite{Mannel:2023yqf}.

In order to compute such quantities one can first integrate analytically the distribution~\eqref{eq:d3Gamma} in $\hat{E}_\ell$ over the whole allowed kinematical range.
For this we need to use the theta functions, which solved for $\hat{E}_\ell$ give
\begin{equation}
\theta(\hat{q}^2) \theta\bigl(-4\hat{E}_\ell^2 + 4\hat{q}_0 \hat{E}_\ell-\hat{q}^2\bigr) = \theta(\hat{q}^2) \theta(\hat{u}_+ - \hat{u}) \theta(\hat{E}_\ell^{\rm max} - \hat{E}_\ell)\theta(\hat{E}_\ell - \hat{E}_\ell^{\rm min})\,,
\end{equation}
with
\begin{equation}
\hat{E}_\ell^{\rm min} = \frac{\hat{q}_0 - \sqrt{\hat{q}_0^2 - \hat{q}^2}}{2}\,, \qquad \hat{E}_\ell^{\rm max} = \frac{\hat{q}_0 + \sqrt{\hat{q}_0^2 - \hat{q}^2}}{2}\,, \qquad \hat{u}_+ = (1-\sqrt{\hat{q}^2})^2 -\rho\,.
\end{equation}
The double differential spectrum is hence simply given by
\begin{equation}
\frac{d^2\Gamma}{d\hat{q}^2 d\hat{u}} = 16 \Gamma_0\theta(\hat{q}^2)\theta(\hat{u}_+-\hat{u})\sqrt{\hat{q}_0^2-\hat{q}^2}\Bigl[3\hat{q}^2 W_1 +(\hat{q}_0^2-\hat{q}^2) W_2 \Bigr]\,.
\end{equation}
The integration over $\hat{u}$ is also straightforward given the form~\eqref{eq:Wstruct} of the structure functions at tree level.
There is a subtlety coming from the fact that the derivatives of the Dirac delta acting on $\theta(\hat{u}_+-\hat{u})$ generate other Dirac deltas (and their derivatives) with argument $\hat{u}_+ = (1-\sqrt{\hat{q}^2})^2-\rho$.
Therefore to look at these terms it is useful to perform a change of variable 
\begin{equation}
\hat{q}^2 = (1-\sqrt{\rho+\hat{u}_+})^2\,, \qquad \biggl|\frac{d\hat{q}^2}{d\hat{u}_+}\biggr| =  \frac{1-\sqrt{\rho+\hat{u}_+}}{\sqrt{\rho+\hat{u}_+}}\,,
\end{equation}
and to simplify the expression using
\begin{equation}
f(z) \delta^{(n)}(z-z_0) = \sum_{k=0}^n (-1)^{n-k} {n \choose k} f^{(n-k)}(z_0) \delta^{(k)}(z-z_0)\,,
\end{equation}
which is easily demonstrated by induction.
The problem is that, starting at order $\mathcal{O}(\lb^3)$, the functions $f(\hat{u}_+)$ in front of the delta function terms $f(\hat{u}_+)\delta^{(n)}(\hat{u}_+)$ have singularities of the form $\mathcal{O}(\hat{u}_+^{-n-1/2})$. These singularities are then cancelled after integration by singularities of the spectrum at the endpoint $\hat{q}^2 \to \hat{q}^2_{\rm max} = (1-\sqrt{\rho})^2$ which is equivalent to $\hat{u}_+ \to 0$. Therefore we need to regulate both terms with the following substitutions\footnote{we choose $\epsilon^2$ as a regulator for convenience so that the singularities are in the form of negative integer powers of $\epsilon$} 
\begin{equation}
\delta^{(n)}(\hat{u}_+) \to \delta^{(n)}(\hat{u}_+-\epsilon^2)\,, \qquad \theta((1-\sqrt{\rho})^2-\hat{q}^2) \to \theta((1-\sqrt{\rho + \epsilon^2})^2-\hat{q}^2) \,,
\end{equation}
before integrating.
After the integration is performed we can finally take the limit $\epsilon \to 0$ and check the cancellation of all singularities.

As a last technicality, the integrals over the spectrum part with no delta functions are much easier to perform with the following change of variable~\cite{Fael:2024gyw}
\begin{equation}
\hat{q}^2 = \biggl(1-\frac{\sqrt{\rho}}{t} \biggr) (1-\sqrt{\rho}  t)\,,
\end{equation}
where the interval $\hat{q}^2 \in [0,(1-\sqrt{\rho})^2]$ is mapped into $t \in [\sqrt{\rho},1]$.
For this reason we have the results for the linear $\hat{q}^2$ moments with a lower cut $\hat{q}^2_{\rm cut}$ as functions of a lower cut on $t$
\begin{equation}
t_{\rm cut} = \frac{1+\rho-\hat{q}^2_{\rm cut}-\sqrt{(1+\rho-\hat{q}^2_{\rm cut})^2-4\rho}}{2\sqrt{\rho}}\,.
\end{equation}
The results of the linear moments, explicitly defined as
\begin{equation}
\mathcal{Q}_n(\hat{q}^2_{\rm cut}) \equiv \frac{1}{\Gamma_0} \int_{\hat{q}^2_{\rm cut}}^{(1-\sqrt{\rho})^2}d \hat{q}^2 (\hat{q}^2)^n \frac{d\Gamma}{d\hat{q}^2}\,,
\end{equation}
are given in the \texttt{Mathematica} ancillary files \texttt{Q$n$.m}, for $n=0,1,2,3$, in the folder \texttt{Resultsq2Mom/}.
They agree up to $\mathcal{O}(\lb^4)$ with the previous determination~\cite{Mannel:2023yqf}, while they differ at order $\mathcal{O}(\lb^5)$ by a term proportional to~\eqref{eq:mb5diff}.

\subsection{Central Moments}
From the moments $\mathcal{M}_n =\{\mathcal{L}_n$, $\mathcal{H}_n,\mathcal{Q}_n\}$ we can build normalized moments $\hat{\mathcal{M}}_n \equiv \mathcal{M}_n/\mathcal{M}_0$, such that the $1/\Gamma_0$ prefactor drops out
and they can be linked to experimental observables independent on the CKM input.
We define the first normalized moments by restoring the proper powers of $m_b$
\begin{equation}
\label{eq:firstmom}
    L_1 = m_b \hat{\mathcal{L}}_1\,, \qquad\qquad H_1 = m_b^2 \hat{\mathcal{H}}_1\,, \qquad\qquad Q_1 = m_b^2 \hat{\mathcal{Q}}_1\,.
\end{equation}
Furthermore, to reduce correlations among different higher moments, it is convenient to study central moments, which are obtained as linear combinations of the normalized moments
\begin{align}
\label{eq:centralmoms}
    L_n &= m_b^n\sum_{k=0}^n (-1)^{n-k} {n \choose k} \hat{\mathcal{L}}_k \hat{\mathcal{L}}_1^{n-k}\,,\qquad\qquad\;\; n\geq 2\,,\nonumber\\
    H_n &= m_b^{2n}\sum_{k=0}^n (-1)^{n-k} {n \choose k} \hat{\mathcal{H}}_k \hat{\mathcal{H}}_1^{n-k}\,,\qquad\qquad n\geq 2\,,\nonumber\\
    Q_n &= m_b^{2n}\sum_{k=0}^n (-1)^{n-k} {n \choose k} \hat{\mathcal{Q}}_k \hat{\mathcal{Q}}_1^{n-k}\,,\qquad\qquad n\geq 2\,,
\end{align}
where we also restored the factors of $m_b$ in each observable. 
The analytic expressions for~\eqref{eq:firstmom}--\eqref{eq:centralmoms} are re-expanded in $\lb$ up to $\mathcal{O}(\lb^5)$ and the results are given in the ancillary files folder \texttt{ResultsCentralMom/} in \texttt{Mathematica} format.
At this stage we should anticipate that the dimensionful scaling in factors of $m_b$ is counteracted by large cancellations in the combinatorial sums, as it will become evident from the phenomenological analysis in Section~\ref{sec:num}.
This, for large $n$, can make such observables quite unstable and sensitive to higher orders and perturbative corrections, as observed in the third central $q^2$ moment~\cite{Finauri:2023kte,Mannel:2023yqf} or third central hadronic mass moment~\cite{Fael:2022frj}.

\section{Numerical Analysis}
\label{sec:num}
We now turn to the numerical evaluation of the moments of the kinematic distribution for $\bar{B} \to X_c \ell \bar{\nu}_\ell$ computed at tree level in the previous section up to order $\mathcal{O}(\lb^5)$ in the HQE.
The input parameters $m_b$, $m_c$, $\mu_\pi^2$, $\mu_G^2$, $\rho_D^3$ and $\rho^3_{LS}$ are extracted from a combined fit to available semileptonic data and lattice results~\cite{Bordone:2021oof,Finauri:2023kte}, as their numerical values is crucial for the inclusive determination of $|V_{cb}|$.
The $b$-quark mass and the HQE parameters are defined in the kinetic scheme\footnote{The expressions in this work are derived in the on-shell scheme. However the difference between the on-shell and kinetic scheme is formally of order $\mathcal{O}(\as)$, allowing us to choose either scheme for the input values of the HQE parameters. We choose the kinetic scheme to be as close as possible to the global fit analyses. In this case the value of $\rho^3_D$ is slightly more than double its size in the on-shell scheme, therefore inducing bigger values for the higher order parameters determined through the LLSA. We consider this choice to be more conservative, as employing the values in the on-shell scheme would have resulted in smaller higher power corrections, but with similar qualitative conclusions.}, while for the charm mass $m_c$ we employ the $\overline{\text{MS}}$ scheme at the renormalization scale of 2 GeV.
The fit results~\cite{Finauri:2023kte} are
\begin{equation}
\label{eq:inputs}
\begin{split}
    m_b &= 4.573 \pm 0.012~\text{GeV}\,,\\
    \mu_\pi^2 &= 0.454\pm 0.043~\text{GeV}^2 \,,\\
    \rho_D^3 &= 0.176 \pm 0.019~\text{GeV}^3\,,
\end{split}\qquad\qquad
\begin{split}
    m_c &= 1.090 \pm 0.010~\text{GeV}\,,\\
    \mu_G^2 &= 0.288 \pm 0.049~\text{GeV}^2\,,\\
    \rho_{LS}^3 &= -0.113 \pm 0.090~\text{GeV}^3\,,
\end{split}
\end{equation}
with the full correlation matrix reported in Table 4 of Ref.~\cite{Finauri:2023kte}. For the $\bar{B}$ meson mass we use $m_B = 5.27972~\text{GeV}$~\cite{ParticleDataGroup:2024cfk} neglecting its small uncertainty.

For the higher order parameters, appearing at $\mathcal{O}(\lb^4)$ and $\mathcal{O}(\lb^5)$, a full extraction from data is not currently possible, as their number grows significantly.
However estimates of the signs and magnitudes of the non-perturbative parameters can be obtained through the LLSA~\cite{Heinonen:2014dxa}.
This approximation allows to write higher order HQE parameters in terms of the lower order ones, $\mu^2_\pi$ and $\mu_G^2$, and two new parameters $\epsilon_{1/2}$ and $\epsilon_{3/2}$.
The parameters $\epsilon_j$ stand for the excitation energies of two degenerate doublet states with angular momentum $\ell=1$ of a fictitious meson made by an hypothetical heavy quark and a light anti-quark, where $j$ is the spin of the light degrees of freedom~\cite{Heinonen:2014dxa}.
In~\cite{Gambino:2016jkc} a global fit including $\mathcal{O}(\lb^5)$ power corrections was performed, including prior distributions for the $\mathcal{O}(\lb^4)$ and $\mathcal{O}(\lb^5)$ HQE parameters coming from the LLSA~\cite{Mannel:2010wj,Heinonen:2014dxa}.
As a result the determined values of the parameters were deviating by the LLSA predictions only by small amounts.

Here we use the LLSA prediction for $\rho_D^3$ and $\rho_{LS}^3$
\begin{align}
    \rho_D^3 &= \frac{1}{3}\epsilon_{1/2}(\mu_\pi^2-\mu_G^2) +\frac{1}{3}\epsilon_{3/2}(2\mu_\pi^2 +\mu_G^2) \,,\nonumber\\
    \rho_{LS}^3 &= \frac{2}{3} \epsilon_{1/2}(\mu_\pi^2 - \mu_G^2) - \frac{1}{3} \epsilon_{3/2}(2\mu_\pi^2 +\mu_G^2)\,,
\end{align}
to determine $\epsilon_{1/2}$ and $\epsilon_{3/2}$ from our inputs~\eqref{eq:inputs}, finding
\begin{equation}
    \epsilon_{1/2} = 379.5~\text{MeV}\,, \qquad \epsilon_{3/2} = 388.8~\text{MeV}\,.
\end{equation}

The LLSA analytic expressions for the dimension 7 and dimension 8 HQE parameters are listed in Appendices~\ref{sec:appdim7} and~\ref{sec:appdim8} respectively.
The accuracy of the LLSA approximation is in general hard to estimate. A toy model examined in~\cite{Heinonen:2014dxa} shows a relative uncertainty of roughly 64\%.
Here we will hence assign a relative error to the parameters determined through the LLSA of 60\% with zero correlation, to be on the conservative side.
Since seven of the dimension 8 HQE parameters have a vanishing LLSA prediction, we set the minimum absolute error to $0.01~\text{GeV}^5 \simeq (400~\text{MeV})^5 \sim \LamQCD^5$.
The LLSA numerical results for the HQE parameters at $\mathcal{O}(\lb^4)$, in GeV$^4$, are 
\begin{align}
    \begin{split}
        m_1 &= 0.115\pm 0.069\,, \\
        m_4 &= 0.36 \pm 0.21\,, \\
        m_7 &= -0.35 \pm 0.21\,,
    \end{split}\qquad
    \begin{split}
        m_2 &= -0.068 \pm 0.041\,, \\
        m_5 &= 0.044 \pm 0.027\,, \\
        m_8 &= -1.05 \pm 0.63\,,
    \end{split}\qquad
    \begin{split}
        m_3 &= -0.055\pm 0.033\,, \\
        m_6 &= 0.055 \pm 0.033\,, \\
        m_9 &= -0.35 \pm 0.21\,.
    \end{split}
\end{align}
At $\mathcal{O}(\lb^5)$, in GeV$^5$, we instead have
\begin{equation}
    \begin{split}
        r_1 &= 0.026\pm 0.016\,, \\
        r_4 &= -0.032 \pm 0.019\,, \\
        r_7 &= 0 \pm 0.01\,, \\
        r_{10} &= 0.051 \pm 0.030\,, \\
        r_{13} &= -0.039 \pm 0.024\,, \\
        r_{16} &= 0 \pm 0.01\,,
    \end{split}\qquad
    \begin{split}
        r_2 &= -0.080 \pm 0.048\,, \\
        r_5 &= 0 \pm 0.01\,, \\
        r_8 &= -0.029 \pm 0.018\,,\\
        r_{11} &= 0.005 \pm 0.010\,, \\
        r_{14} &= 0.029 \pm 0.017\,, \\
        r_{17} &= 0 \pm 0.01\,,
    \end{split}\qquad
    \begin{split}
        r_3 &= -0.021\pm 0.013\,, \\
        r_6 &= 0 \pm 0.01\,, \\
        r_9 &= 0.051 \pm 0.031\,,\\
        r_{12} &= 0.006 \pm 0.010\,, \\
        r_{15} &= 0 \pm 0.01\,, \\
        r_{18} &= 0 \pm 0.01\,.
    \end{split}
\end{equation}

\begin{figure}
    \centering
    \includegraphics[width=\textwidth]{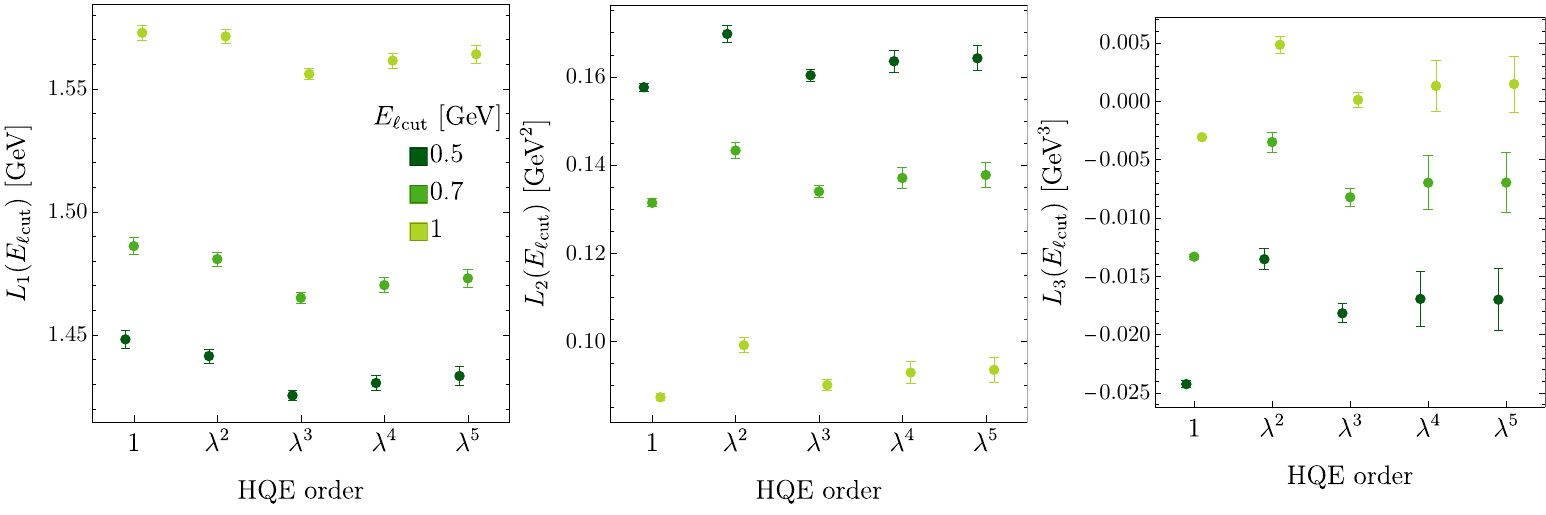}\\
    \includegraphics[width=\textwidth]{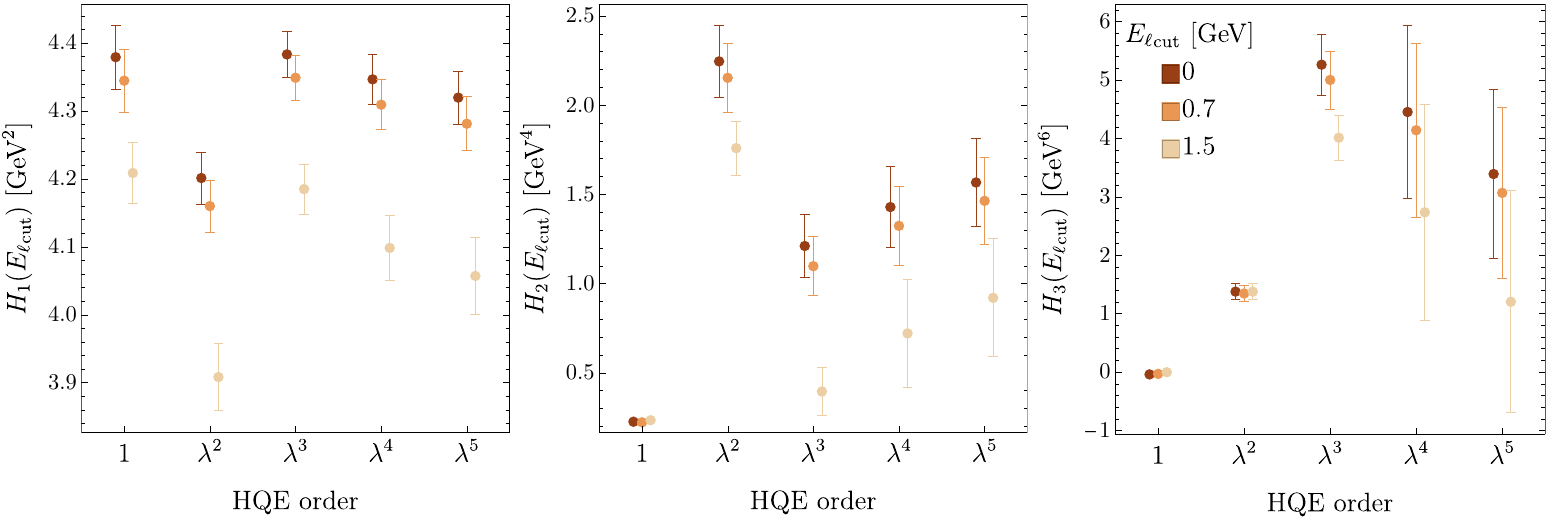}\\
    \includegraphics[width=\textwidth]{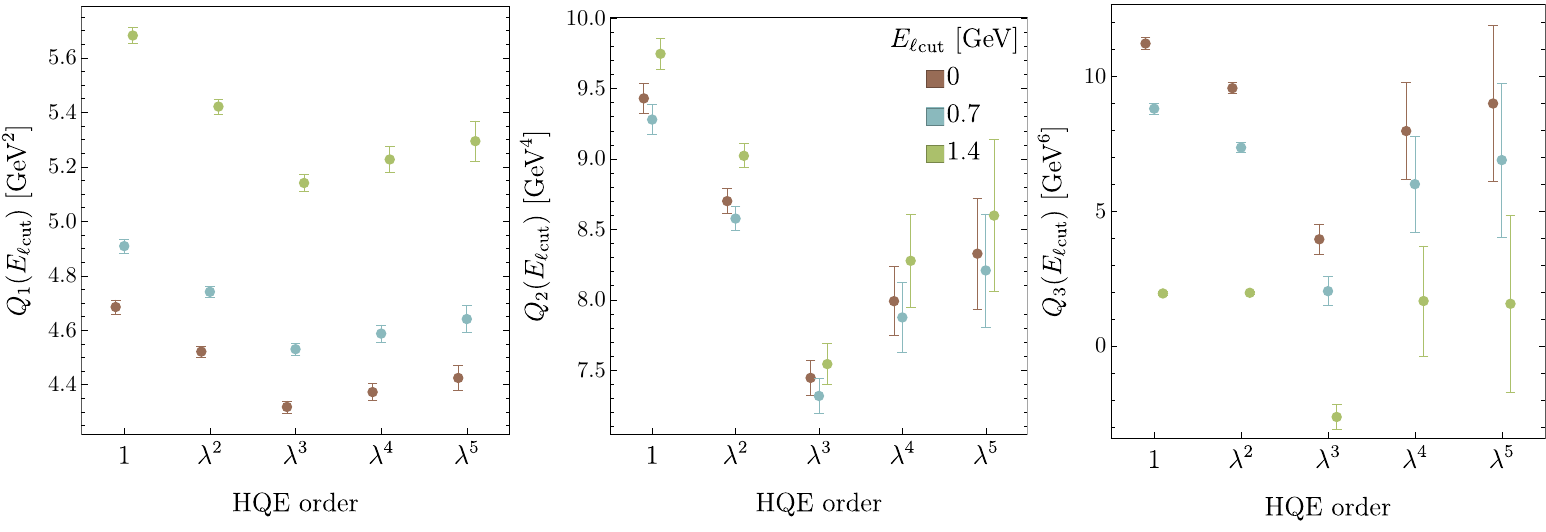}
    \caption{\small First (left), second central (middle) and third central (right) moments in the lepton energy (top), hadronic invariant mass (middle) and $q^2$ (bottom). All moments computed with three different a lower cuts ${E_{\ell}}_{\rm cut}$ on the lepton energy. The moments are displayed as functions of the order at which the HQE is truncated. Uncertainties are obtained by gaussian error propagation from the inputs.}
    \label{fig:LHQplot}
\end{figure}

We show in Figure~\ref{fig:LHQplot} the results for $L_n({E_{\ell}}_{\rm cut})$, $H_n({E_{\ell}}_{\rm cut})$ and $Q_n({E_{\ell}}_{\rm cut})$ as a function of the order in the HQE (for example the point at $\lb^3$ corresponds to the theoretical prediction including corrections up to $\mathcal{O}(\lb^3)$). The different colours denote different lepton energy cuts.
It is important to notice that we do not include a theoretical uncertainty accounting for missing higher orders and perturbative corrections, as our goal here is not to make testable theoretical predictions, but to investigate the convergence of the tree level HQE.
In fact, this analysis is meant to help in estimating the theoretical uncertainties due to missing higher orders in the HQE, in particular for fits to semileptonic moments where the truncation usually happens at $\mathcal{O}(\lb^3)$. We will elaborate more on this point in Section~\ref{sec:thunc}.
It is hence expected that in Figure~\ref{fig:LHQplot} the uncertainties grow when including higher order corrections, as the new non-perturbative parameters come with larger errors.
We notice in general that, for standard values of the lepton energy cut ${E_{\ell}}_{\rm cut} <1.5~\text{GeV}$, the behaviour of the HQE is independent of the cut.
Also we see that the dimension 8 contributions seem to be sufficiently under control for all the observables, despite the sizeable increase in number of independent parameters.
On the other hand the shift in the central value for the hadronic and $q^2$ moments is sensible when adding the $\mathcal{O}(\lb^4)$ corrections.
In particular for the third $q^2$ moment truncating the expansion at $\mathcal{O}(\lb^3)$ seems to introduce a non negligible bias.
Nevertheless we should of course keep in mind that the values of the HQE parameters $\mu^2_\pi$, $\mu^2_G$, $\rho_D^3$, $\rho_{LS}^3$ in the plot are extracted from experimental data using theoretical predictions truncated at $\mathcal{O}(\lb^3)$.
Therefore it is likely that their numerical value is partly taking into account the higher order corrections.

\begin{figure}
    \centering
    \includegraphics[width=\textwidth]{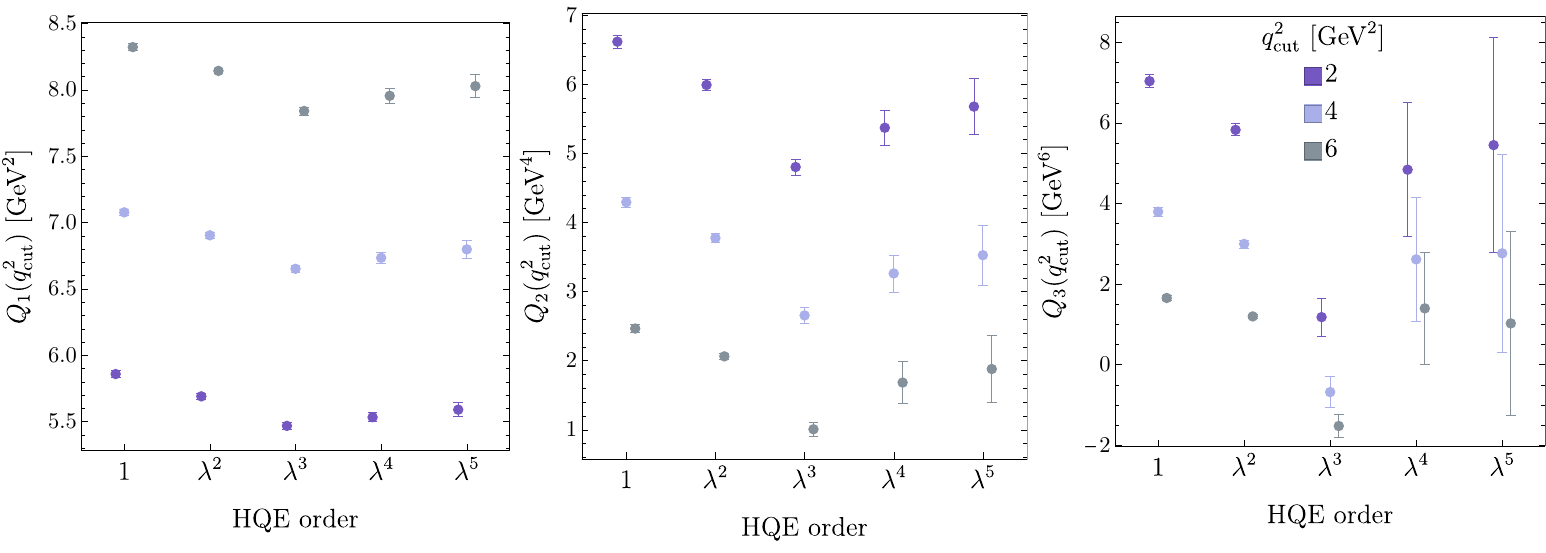}
    \caption{\small First (left), second central (middle) and third central (right) $q^2$ moments with three different lower cuts in $q^2$. The moments are displayed as functions of the order at which the HQE is truncated. Uncertainties are obtained by gaussian error propagation from the inputs.}
    \label{fig:Qplot}
\end{figure}

In Figure~\ref{fig:Qplot} we show similar plots but for the $q^2$ moments with a lower cut $q^2_{\rm cut}$ in $q^2$.
No substantial differences are observed with respect to the $q^2$ moments with a cut on the lepton energy and the same comments apply.

\subsection{Convergence of the $\rho$ Expansion}
In the OPE, the charm quark is treated as a heavy quark, being integrated out simultaneously with the $b$.
The analytic expressions for the structure functions, and hence the moments, are then functions of the $\mathcal{O}(1)$ variable $\rho = m_c^2/m_b^2$.
However given the numerical value $\rho \simeq 0.05$ one would expect that a series in $\rho$ could give a reasonable approximation for the full expressions of the moments.
In this section we investigate the behaviour of the expansion in $\rho$ for each order of the HQE separately.

The phase space integral exhibits an infrared sensitivity to the charm mass, first at order $\mathcal{O}(\lb^3)$ through a purely logarithmic term $\ln \rho$~\cite{Breidenbach:2008ua}.
In particular, at order $\mathcal{O}(\lb^5)$, terms proportional to $1/\rho$ arise, often called ``intrinsic charm'' contributions~\cite{Bigi:2009ym,Mannel:2010wj, Mannel:2023yqf}.
It is argued in the literature that terms of the form $\mathcal{O}(\LamQCD^5/(m_c^2 m_b^3))$ could indeed provide the bulk of the dimensions 8 contribution, and even be counted formally as $\mathcal{O}(\LamQCD^4/m_b^4)$, in light of the actual charm mass value $m_c \sim \sqrt{\LamQCD m_b}$.
It was shown in~\cite{Mannel:2023yqf} that these contributions for the third central $q^2$ moment were anomalously large (at least twice the leading power term), but almost completely cancel against the rest of $\mathcal{O}(\lb^5)$ terms. 
We confirm this qualitative observation, but with the absolute magnitude of the intrinsic charm contributions strongly depending on the input values for the LLSA.

\begin{figure}
    \centering
    \includegraphics[width=\linewidth]{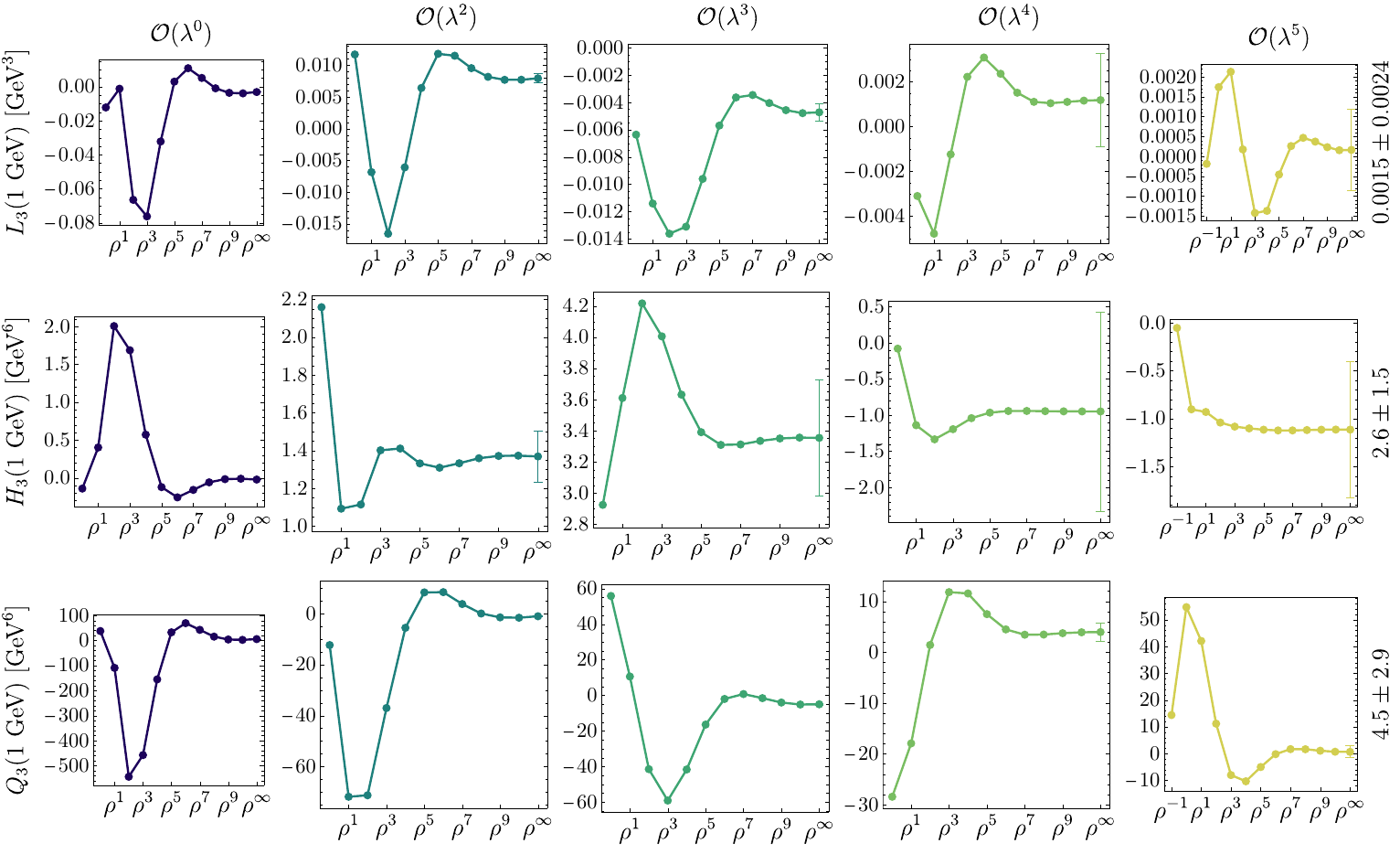}
    \caption{\small Third central moment in the lepton energy (top), hadronic invariant mass (middle) and $q^2$ (bottom) at ${E_{\ell}}_{\rm cut}=1~\text{GeV}$. Each column contains only terms of the corresponding $\mathcal{O}(\lb^k)$ labelled at the top. The number displayed on the right corresponds to the final sum of all orders, with respective uncertainty. Each panel shows the $\rho$ expansion of the respective term, where the last point (at $\rho^\infty$) corresponds to the unexpanded full term including its uncertainty.}
    \label{fig:rho3}
\end{figure}

To better understand, and test, the quality of the $\rho$ expansion, we plot the result of expanding each term in the OPE separately in $\rho$ up to a given order.
Notice that due to the presence of logarithms of $\rho$ this is not a Taylor expansion, and we will treat $\ln \rho$ as $\mathcal{O}(1)$ constants.
In particular in Figure~\ref{fig:rho3} we show the third central moments $L_3$, $H_3$ and $Q_3$ at ${E_{\ell}}_{\rm cut}=1~\text{GeV}$ where each column stands for the isolated $\mathcal{O}(\lb^k)$ term of the OPE.
The $x$-axes show the order at which the series in $\rho$ is truncated, starting at $\rho^0$ for the terms up to $\mathcal{O}(\lb^4)$, and at $\rho^{-1}$ for the terms $\mathcal{O}(\lb^5)$.
The last point of each plot, at $\rho^\infty$, stands for the unexpanded full expression, where we also indicate the uncertainty.
To further understand the impact of each point on the total prediction, we display at the end of the row on the right the result for the full OPE prediction, which includes the full $\rho$ dependence and up to $\mathcal{O}(\lb^5)$ corrections.
As previously anticipated, the third central moments are rather unstable, and the $\rho$ expansion is almost never a good approximation, unless at least terms up to $\mathcal{O}(\rho^6)$ are included.
The case of the third central $q^2$ moment stands out as particularly unstable (as already noticed in Figures~\ref{fig:LHQplot}--\ref{fig:Qplot}) where the simple leading power term if expanded up to $\mathcal{O}(\rho^2)$  gives $Q_3(1~\text{GeV}) \approx -544~\text{GeV}^6$, to be compared with the full result of $4.5\pm 2.9~\text{GeV}^6$.

\begin{figure}
    \centering
    \includegraphics[width=\linewidth]{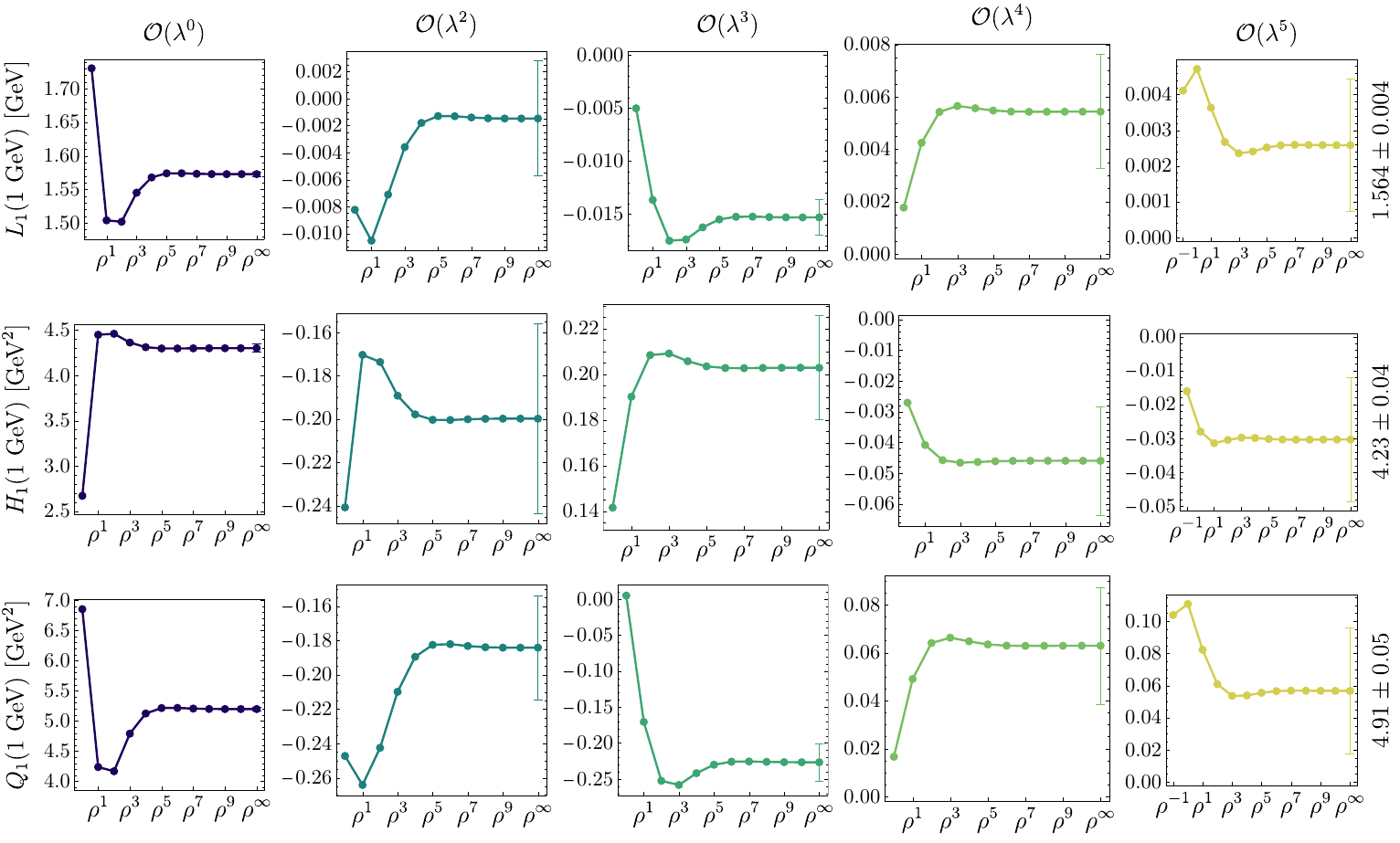}
    \caption{\small First moment in the lepton energy (top), hadronic invariant mass (middle) and $q^2$ (bottom) at ${E_{\ell}}_{\rm cut}=1~\text{GeV}$. See caption of Figure~\ref{fig:rho3} for further details.}
    \label{fig:rho1}
\end{figure}
\begin{figure}
    \centering
    \includegraphics[width=\linewidth]{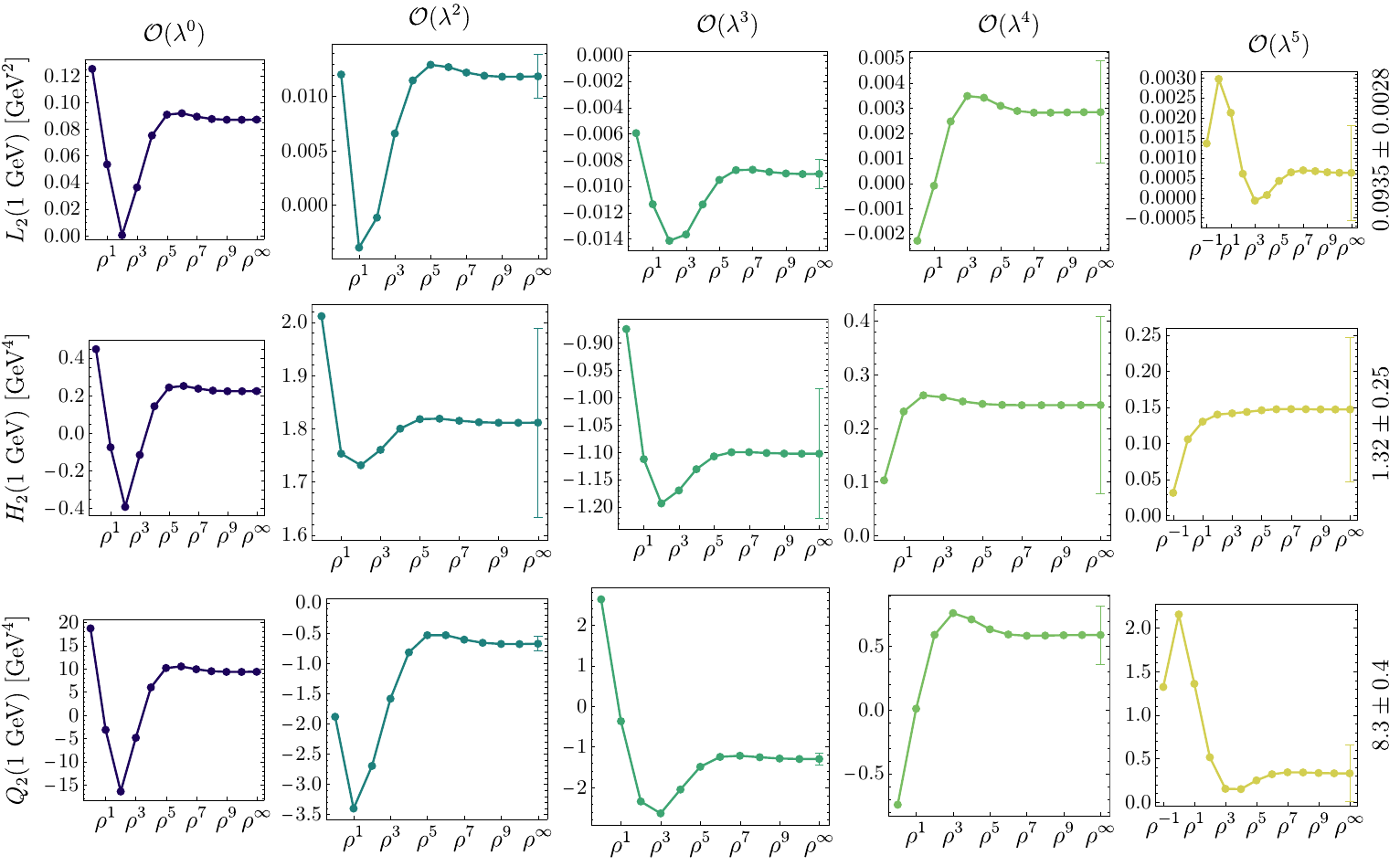}
    \caption{\small Second central moment in the lepton energy (top), hadronic invariant mass (middle) and $q^2$ (bottom) at ${E_{\ell}}_{\rm cut}=1~\text{GeV}$. See caption of Figure~\ref{fig:rho3} for further details.}
    \label{fig:rho2}
\end{figure}
For completeness we also report the analogous of Figure~\ref{fig:rho3}, but for the first (Figure~\ref{fig:rho1}) and second central moments (Figure~\ref{fig:rho2}).
In such cases one observes a more robust convergence of the $\rho$ expansion, but where nevertheless the first order approximation is rarely satisfactory.
We suspect that the convergence of the $\rho$ expansion is spoiled by the potentially large logarithmic terms $\ln \rho$.
To improve the convergence of the expansion one would probably need to implement a two step matching (by considering $m_b \gg m_c \gg \LamQCD$), which would allow to resum the leading logarithmic contributions~\cite{Bauer:1996ma,Breidenbach:2008ua}.

\subsection{Data on $q^2$ Moments with Lepton Energy Cut}
\label{sec:CLEO}
In this section we would like to compare our tree level predictions for $Q_1({E_\ell}_{\rm cut})$ and $Q_2({E_\ell}_{\rm cut})$ with aforementioned measurements from the CLEO collaboration~\cite{CLEO:2004bqt}.
The potential interest relies on the fact that, to the best of our knowledge, $q^2$ moments with a lower cut on the lepton energy where never employed in fits for the inclusive extraction of $|V_{cb}|$.
With our knew theoretical predictions up to $\mathcal{O}(\lb^5)$, supplemented with the easily implementable first perturbative corrections $\mathcal{O}(\as)$, they could enter global fits as new interesting observables.
We also imagine that a combined extraction of $E_\ell$, $m_X^2$ and $q^2$ moments with a cut on the lepton energy, including experimental correlations, is feasible with the current Belle and Belle II datasets.

The experimental results from the CLEO collaboration\footnote{We summed in quadrature statistic and systematic uncertainties.}~\cite{CLEO:2004bqt}
\begin{align}
\langle q^2 \rangle(E_\ell > 1~\text{GeV}) &= 4.89 \pm 0.14~\text{GeV}^2\,, \nonumber\\
\langle q^2 \rangle(E_\ell > 1.5~\text{GeV}) &= 5.29 \pm 0.12~\text{GeV}^2\,, \nonumber\\
\langle (q^2-\langle q^2\rangle)^2 \rangle(E_\ell > 1~\text{GeV}) &= 2.852 \pm 0.047~\text{GeV}^4\,, \nonumber\\
\langle (q^2-\langle q^2\rangle)^2 \rangle(E_\ell > 1.5~\text{GeV}) &= 2.879 \pm 0.050~\text{GeV}^4\,,
\end{align}
raise interesting questions when compared to our theoretical predictions
\begin{align}
\label{eq:thpredq2Ecut}
    Q_1(1~\text{GeV}) &= 4.91 \pm 0.05 ~\text{GeV}^2\,,\nonumber\\
    Q_1(1.5~\text{GeV}) &= 5.35\pm 0.09~\text{GeV}^2\,,\nonumber\\
    Q_2(1~\text{GeV}) &=8.3 \pm 0.4~\text{GeV}^4\,,\nonumber\\
    Q_2(1.5~\text{GeV}) &= 8.6 \pm 0.6~\text{GeV}^4\,.
\end{align}
The predictions~\eqref{eq:thpredq2Ecut} do not include perturbative corrections, but given the known size of perturbative corrections in the first two $q^2$ moments with lower cut on $q^2$~\cite{Finauri:2023kte,Fael:2024gyw}, we do not expect more than a few percent deviation from~\eqref{eq:thpredq2Ecut}.
We postpone to a future work the implementation of the $\mathcal{O}(\as)$ contributions to these observables.

Further elements to consider in this discussion are the Belle II\footnote{The Belle collaboration also performed a similar analysis~\cite{Belle:2021idw} but the lowest value for $q^2_{\rm cut}$ they considered is 3~GeV$^2$.}~\cite{Belle-II:2022evt} $q^2$ moments measurements with a lower cut in $q^2$.
At the lowest available value of the cut~\cite{Belle-II:2022evt}
\begin{align}
    \langle q^2 \rangle(q^2>1.5~\text{GeV}^2) &= 5.16\pm 0.11~\text{GeV}^2\,,\nonumber\\
    \langle (q^2-\langle q^2\rangle)^2 \rangle(q^2>1.5~\text{GeV}^2) &= 5.97\pm 0.24~\text{GeV}^4\,,
\end{align}
we expect these measurements to be not too far from the $q^2$ moments measured with a lower cut in the lepton energy at the lowest available value of the cut\footnote{The lowest available cut in the lepton energy is 1~GeV, which is approximately in the middle of the allowed phase space in the lepton energy. However we checked that for $0<{E_\ell}_{\rm cut} <1~\text{GeV}$ the functions $Q_1({E_\ell}_{\rm cut})$ and $Q_2({E_\ell}_{\rm cut})$ do not vary too drastically.}.
This is because for vanishing cuts the two observables coincide.

The second central $q^2$ moment measured by CLEO, given the quoted uncertainty, stands out as very far from both our theoretical prediction and the Belle II data obtained with a lower cut $q^2_{\rm cut} = 1.5~\text{GeV}^2$.
This makes an update of such measurement quite appealing, as no conclusions can be drawn from the limited available data.
To suppress possible systematic effects we can look at the ratio between the value at ${E_\ell}_{\rm cut} =1.5~\text{GeV}$ and ${E_\ell}_{\rm cut} = 1~\text{GeV}$
\begin{equation}
    \frac{\langle (q^2-\langle q^2\rangle)^2 \rangle(E_\ell > 1.5~\text{GeV})}{\langle (q^2-\langle q^2\rangle)^2 \rangle(E_\ell > 1~\text{GeV})} = 1.009\pm 0.024\,,
\end{equation}
which is compatible with out theoretical expectation
\begin{equation}
    \frac{Q_2(1.5~\text{GeV})}{Q_2(1~\text{GeV})} = 1.027\pm 0.020\,.
\end{equation}

\subsection{Theoretical Uncertainties from Higher Orders}
\label{sec:thunc}
\begin{figure}
    \centering
    \includegraphics[width=\linewidth]{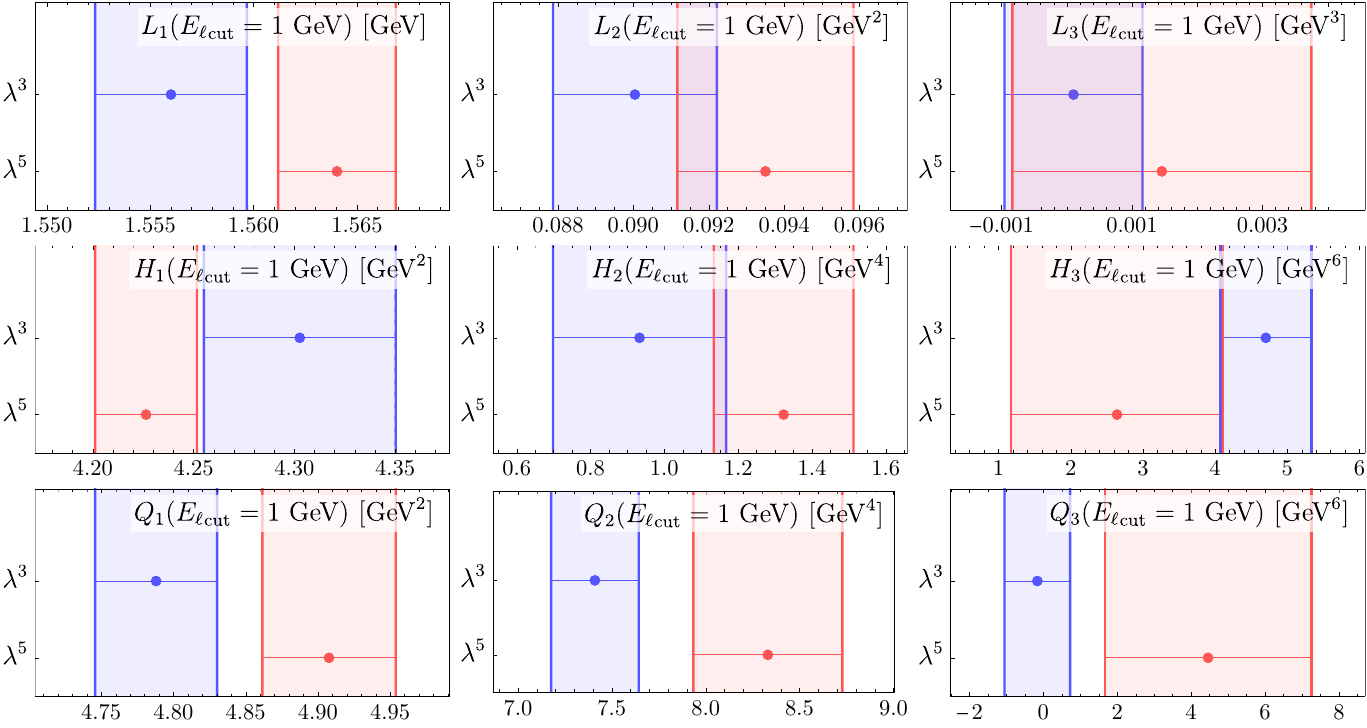}
    \caption{\small Plots for the comparison of the $\mathcal{O}(\lb^3)$ predictions with theory uncertainty estimates (blue), against the $\mathcal{O}(\lb^5)$ predictions with uncertainty coming only from the $m_i$ and $r_i$ HQE parameters.  All moments are computed with ${E_\ell}_{\rm cut} = 1~\text{GeV}$.}
    \label{fig:gridThUnc}
\end{figure}

In this last section we want to investigate whether the theoretical uncertainties assigned to the predictions of the moments in~\cite{Finauri:2023kte}, including up to $\mathcal{O}(\lb^3)$ terms, 
are appropriately accounting for the missing $\mathcal{O}(\lb^4)$ and $\mathcal{O}(\lb^5)$ terms computed in this work.
For doing this we compute the moments cutting the HQE at order $\mathcal{O}(\lb^3)$, and assigning the theoretical uncertainty due to missing higher orders by setting the following uncorrelated absolute errors on the input parameters
\begin{equation}
\begin{split}
    \sigma^{\rm th}_{m_b} &= 0.004~\text{GeV}\,,\\
    \sigma^{\rm th}_{\mu_\pi^2} &= 0.030~\text{GeV}^2 \,,\\
    \sigma^{\rm th}_{\rho_D^3} &= 0.032~\text{GeV}^3\,,
\end{split}\qquad\qquad
\begin{split}
    \sigma^{\rm th}_{m_c} &= 0.004~\text{GeV}\,,\\
    \sigma^{\rm th}_{\mu_G^2} &= 0.020~\text{GeV}^2\,,\\
    \sigma^{\rm th}_{\rho_{LS}^3} &= 0.018~\text{GeV}^3\,.
\end{split}
\end{equation}
Notice that we will not include the uncertainties coming from the input parameters~\eqref{eq:inputs} as we are only interested in the assigned theoretical uncertainty.
For the full $\mathcal{O}(\lb^5)$ prediction we also do not include the uncertainty coming from the input parameters~\eqref{eq:inputs}, but only the one coming from the variation of the $\mathcal{O}(\lb^4)$ and $\mathcal{O}(\lb^5)$ HQE parameters, as that one should give a reasonable estimate of the possible range of values coming from HQE orders higher than $\mathcal{O}(\lb^3)$.
We show our results in Figure~\ref{fig:gridThUnc} for ${E_\ell}_{\rm cut} = 1~\text{GeV}$ (for $q^2$ moments with a cut in $q^2$ the situation is qualitatively analogous to what is shown for $Q_i({E_\ell}_{\rm cut})$).
We notice that $q^2$ moments are again the more problematic ones, where the theory uncertainty assigned to the truncated results seems to always underestimate the contribution from higher power corrections.
It is therefore recommended to either increase the theoretical uncertainties in such observables, or perform a combined fit with at least the $\mathcal{O}(\lb^4)$ tree level corrections, along the lines of~\cite{Gambino:2016jkc,Bernlochner:2022ucr}.
In general we still want to emphasize that the convergence of the HQE, after having performed a new fit including higher order corrections, can substantially improve through the change in the numerical values of the lower dimensional HQE parameters.

\section{Summary}
\label{sec:summ}
In this work we have investigated the behaviour of the HQE (at tree level) up to $\mathcal{O}(\lb^5)$, by updating the analytic results for the triple differential decay rate and the $E_\ell$, $m_X^2$ and $q^2$ moments~\cite{Mannel:2010wj,Mannel:2023yqf,Fael:2024fkt}.
Our results are available in \texttt{Mathematica} format in the ancillary files.
Through the LLSA~\cite{Mannel:2010wj,Heinonen:2014dxa} we were able to test the numerical impact of the $\mathcal{O}(\lb^4)$ and $\mathcal{O}(\lb^5)$ terms of the HQE on the first three moments in the lepton energy, hadronic invariant mass and $q^2$.
We found that the third central $q^2$ moment appears as a rather unstable observable against power corrections~\cite{Finauri:2023kte}, requiring a careful treatment of theoretical uncertainties.
In addition we have investigated the quality of the expansion in $m_c \ll m_b$, finding that, without resumming phase space logarithms, the convergence of the expansion is very slow.
This suggests that the counting $m_c \sim \mathcal{O}(m_b)$ is indeed appropriate for inclusive semileptonic $\bar{B}$ decays.

We have derived for the first time analytic expressions for $q^2$ moments with a lower cut on the lepton energy up to order $\mathcal{O}(\lb^5)$.
As these quantities were measured by the CLEO collaboration roughly 20 years ago, we investigate the compatibility of such measurements with our theoretical predictions.
We find good agreement with the first normalized moment, while the second central moment presents a puzzling discrepancy, highlighting the importance of new measurements of these observables by other collaborations.
If included in a global fit, these measurements could improve the precision of $|V_{cb}|$ extracted from inclusive decays.

Often the HQE predictions are cut at $\mathcal{O}(\lb^3)$ in order to perform a full extraction of the non-perturbative parameters from data.
We have compared this case, including the assigned theoretical uncertainty, to the full result up to $\mathcal{O}(\lb^5)$ where we have included only the uncertainty coming from the $\mathcal{O}(\lb^4)$ and $\mathcal{O}(\lb^5)$ parameters.
With this we have concluded that in the case of the $q^2$ moments the assigned theoretical uncertainty is probably to be improved with the knowledge of the tree level higher power corrections.

We leave to a future work the implementation of the analytic results presented in this paper into a global fit for the inclusive determination of $|V_{cb}|$.

\subsubsection*{Acknowledgements}
We would like to thank Paolo Gambino for extensive discussions, and for the careful reading of the manuscript.
We thank Ilija Milutin and Keri Vos for useful correspondence on the details of Ref.~\cite{Mannel:2023yqf}, as well as Martin Jung for helpful discussions.
We are grateful to the Mainz Institute for Theoretical Physics (MITP) of the Cluster of Excellence PRISMA$^+$ (Project ID 390831469), for its hospitality and support during the early stage of this project.
This work is supported by the Italian Ministry of Research (MIUR) under the grant 2022N4W8WR.

\appendix
\section{HQE Parameters}
\label{sec:appHQE}
\subsection{Dimension 3 and 4}
The following matrix elements of dimension 3 operators are exact in QCD
\begin{align}
\langle \bar{B} |\bar{b} \gamma^\mu b |\bar{B}\rangle &= 2m_B v^\mu\,, \nonumber\\
\langle \bar{B} |\bar{b} \gamma^\mu \gamma^5 b |\bar{B}\rangle &= 0\,,
\end{align}
which also hold for the rephased field $b_v$.
The matrix elements are taken between full QCD states.

The operators at dimension 4, $\bar{b}_v (i v\cdot D) b_v$ and $\bar{b}_v i \slashed{D} b_v$, are related to higher-dimensional operators through the equations of motion~\eqref{eq:bveom1}-\eqref{eq:bveom2}. Therefore, there are no HQE parameters of mass dimension 1.
To see this explicitly one can use
\begin{equation}
    i \slashed{D} = (i v\cdot D)\slashed{v} + i \slashed{D}_\perp = (i v\cdot D)\slashed{v} + i \slashed{D}_\perp P_- + P_- i \slashed{D}_\perp\,,
\end{equation}
and use the fact that $iv\cdot D$ and $P_-$ acting on the field $b_v$ give a higher dimensional contribution.

\subsection{Dimension 5 and 6}
At dimension 5 there are two independent HQE parameters defined from
\begin{align}
2m_B \mu_\pi^2 &= -\langle \bar{B}|\bar{b}_v iD_\perp^\mu i{D_\perp}_\mu b_v |\bar{B}\rangle\,,\nonumber\\
2m_B \mu_G^2 &= \frac{1}{2} \langle \bar{B}|\bar{b}_v [iD^\mu_\perp,iD^\nu_\perp](-i\sigma_{\mu\nu}) b_v |\bar{B}\rangle\,.
\end{align}

At dimension 6 there are also two independent HQE parameters
\begin{align}
2m_B \rho_D^3 &= \frac{1}{2}\langle \bar{B}|\bar{b}_v \biggl[i D_\perp^\mu ,\Bigl[i v \cdot D,i D_{\perp\mu}\Bigr]\biggr] b_v |\bar{B}\rangle\,,\nonumber\\
2m_B \rho^3_{LS} &= \frac{1}{2}\langle \bar{B}|\bar{b}_v \biggl\{i D_\perp^\mu ,\Bigl[i v \cdot D,i D^\nu_\perp\Bigr]\biggr\} (-i\sigma_{\mu\nu}) b_v |\bar{B}\rangle\,.
\end{align}

\subsection{Dimension 7}
\label{sec:appdim7}
At dimension 7 there are 9 independent parameters, where one also has to take into account the constraint that, due to $T$ invariance, in the OPE only Hermitian operators can appear~\cite{Mannel:2010wj}.
The full list is~\cite{Mannel:2010wj}
\begin{align}
\label{eq:ord4pars}
2m_B m_1 &= \frac{1}{3}\langle \bar{B} | \bar{b}_v \, i D_\rho i D_\sigma i D_\xi iD_\delta \,b_v | \bar{B} \rangle\,\left(g_\perp^{\rho \sigma}g_\perp^{\xi \delta} + g_\perp^{\rho \xi}g_\perp^{\sigma \delta}+g_\perp^{\rho \delta}g_\perp^{\sigma \xi}\right)\,, \nonumber\\
2m_B m_2 &= \langle \bar{B} | \bar{b}_v \, [ i D_{\perp}^\mu ,  i v\cdot D ] [ i v\cdot D ,  i D_{\perp\mu}] \, b_v | \bar{B} \rangle\,, \nonumber\\
2m_B m_3 &= \langle \bar{B} | \bar{b}_v \, [i D^\mu_\perp ,  i D_\perp^\nu][ iD_{\perp\mu}, iD_{\perp\nu}] \,b_v| \bar{B} \rangle\,, \nonumber\\
2m_B m_4 &= \langle \bar{B} | \bar{b}_v \, \Bigl\{ i D_\perp^\mu , \Bigl[i D_{\perp}^\nu, [ i D_{\perp\nu} , i D_{\perp\mu}]\Bigr]\Bigr\} b_v | \bar{B} \rangle\,, \nonumber\\
2m_B m_5 &= \langle \bar{B} | \bar{b}_v \, [ i D_\perp^\mu  , i v \cdot D][ i v\cdot D ,  i D_\perp^\nu]  (-i \sigma_{\mu \nu})\,b_v | \bar{B} \rangle\,, \nonumber\\
2m_B m_6 &= \langle \bar{B} | \bar{b}_v \, [ i D_\perp^\rho , i D_\perp^\mu ][i D_\perp^\nu , i D_{\perp\rho}]  (-i \sigma_{\mu \nu})\,b_v| \bar{B} \rangle\,, \nonumber\\
2m_B m_7 &= \langle \bar{B} | \bar{b}_v \,\Bigl \{ \{ i D_\perp^\mu, i D_\perp^\rho \} ,[ i D_{\perp\rho} , i D_\perp^\nu] \Bigr\} (-i \sigma_{\mu \nu})\,b_v | \bar{B} \rangle \,,\nonumber\\
2m_B m_8 &= \langle \bar{B} | \bar{b}_v \,\Bigl\{ \{ i D_\perp^\rho ,  i D_{\perp\rho} \}, [i D_\perp^\mu , i D_\perp^\nu]\Bigr\} (-i \sigma_{\mu \nu})\,b_v | \bar{B} \rangle \,,\nonumber\\
2m_B m_9 &= \langle \bar{B} | \bar{b}_v \, \biggl[ i D_\perp^\nu ,\Bigl[ i D_\perp^\rho , [ i D_\perp^\mu ,  i D_{\perp\rho} ]\Bigr]\biggr] (-i \sigma_{\mu \nu})\,b_v | \bar{B} \rangle \,.
\end{align}
The LLSA prediction for the parameters~\eqref{eq:ord4pars} is~\cite{Heinonen:2014dxa}
\begin{equation}
    \begin{split}
    m_1 &= \frac{5}{9}\mu_\pi^4 \,,\\
    m_3 &= -\frac{2}{3}\mu_G^4\,,\\
    m_5 &= -\frac{2}{3}\epsilon_{1/2}^2 (\mu_\pi^2-\mu_G^2) +\frac{\epsilon_{3/2}^2}{3}(2\mu_\pi^2+\mu_G^2)\,,\\
    m_7 &= -\frac{8}{3}\mu_\pi^2 \mu_G^2\,,\\
    m_9 &= \mu_G^4-\frac{10}{3}\mu_\pi^2\mu_G^2\,,
    \end{split}\qquad
    \raisebox{0.7cm}{$\begin{split}
    m_2 &= -\frac{\epsilon_{1/2}^2}{3}(\mu_\pi^2-\mu_G^2) -\frac{\epsilon_{3/2}^2}{3}(2\mu_\pi^2+\mu_G^2) \,,\\
    m_4 &= \mu_G^4 + \frac{4}{3}\mu_\pi^4\,,\\
    m_6 &= \frac{2}{3}\mu_G^4\,,\\
    m_8 &= -8\mu_\pi^2 \mu_G^2\,,\phantom{\frac{8}{3}}
    \end{split}$}
\end{equation}

\subsection{Dimension 8}
\label{sec:appdim8}
At dimension 8 there are 18 independent parameters~\cite{Mannel:2010wj}.
Also in this case the constraint on the hermiticity of the operators has been used; however, the parameters are defined from matrix elements of operators which are not necessarily hermitian.
This means that such parameters, as defined in Ref.~\cite{Mannel:2010wj}, could in principle have an imaginary part.
Here we define the basis of parameters with the same set of operators, but taking the real part of the matrix element, which can be seen as adding the corresponding hermitian conjugate operator to the definition in order to have only hermitian operators
\begin{align}
\label{eq:ord5pars}
2m_B r_1 &= \text{Re}\Bigl\{ \langle \bar{B} | \bar{b}_v \,i  D_\rho\, (i v \cdot D)^3\, i  D^\rho \, b_v | \bar{B} \rangle \Bigr\}\,, \nonumber \\
2m_B r_2 &= \text{Re}\Bigl\{\langle \bar{B} | \bar{b}_v \,i  D_\rho\, (i v \cdot D)\, i  D^\rho\, i  D_\sigma\, i  D^\sigma \, b_v | \bar{B} \rangle \Bigr\}\,, \nonumber\\
2m_B r_3 &= \text{Re}\Bigl\{\langle \bar{B} | \bar{b}_v \,i  D_\rho\, (i v \cdot D)\, i  D_\sigma\, i D^\rho\, i  D^\sigma \, b_v | \bar{B} \rangle \Bigr\}\,, \nonumber \\
2m_B r_4 &= \text{Re}\Bigl\{\langle \bar{B} | \bar{b}_v \,i  D_\rho\, (i v \cdot D)\, i  D_\sigma\, i D^\sigma\, i  D^\rho \, b_v | \bar{B} \rangle \Bigr\}\,, \nonumber \\
2m_B r_5 &= \text{Re}\Bigl\{\langle \bar{B} | \bar{b}_v \,i  D_\rho\, i  D^\rho\,(i v \cdot D)\,  i D_\sigma\, i  D^\sigma \, b_v | \bar{B} \rangle \Bigr\}\,,\nonumber \\
2m_B r_6 &= \text{Re}\Bigl\{\langle \bar{B} | \bar{b}_v \,i  D_\rho\, i  D_\sigma\, (i v \cdot D)\, i D^\sigma\, i  D^\rho \, b_v | \bar{B} \rangle\Bigr\} \,, \nonumber\\
2m_B r_7 &= \text{Re}\Bigl\{\langle \bar{B} | \bar{b}_v \,i  D_\rho\, i  D_\sigma\, (i v \cdot D)\, iD^\rho\, i  D^\sigma \, b_v | \bar{B} \rangle \Bigr\}\,, \nonumber \\
2m_B r_8 &= \text{Re}\Bigl\{\langle \bar{B} | \bar{b}_v \,i  D_\mu \, (i v \cdot D)^3\, i   D_\nu \, (-i \sigma^{\mu \nu })\,b_v | \bar{B} \rangle \Bigr\}\,,\nonumber \\
2m_B r_9 &= \text{Re}\Bigl\{\langle \bar{B} | \bar{b}_v \,i  D_\mu \, (i v \cdot D)\, i D_\nu\, i   D_\rho\, i   D^\rho \,(-i \sigma^{\mu \nu })\, b_v | \bar{B} \rangle \Bigr\}\,,\nonumber \\ 
2m_B r_{10} &= \text{Re}\Bigl\{\langle \bar{B} | \bar{b}_v \,i D_\rho\, (i v \cdot D)\, i D^\rho\, i   D_\mu \, i   D_\nu  \,(-i \sigma^{\mu \nu })\, b_v | \bar{B} \rangle \Bigr\}\,, \nonumber \\
2m_B r_{11} &= \text{Re}\Bigl\{\langle \bar{B} | \bar{b}_v \,i D_\rho\, (i v \cdot D)\, i   D_\mu \, i   D^\rho\, i   D_\nu  \,(-i \sigma^{\mu \nu })\, b_v | \bar{B} \rangle \Bigr\}\,, \nonumber \\
2m_B r_{12} &= \text{Re}\Bigl\{\langle \bar{B} | \bar{b}_v \,i D_\mu \, (i v \cdot D)\, i D_\rho\, i   D_\nu \, i   D^\rho \,(-i \sigma^{\mu \nu })\, b_v | \bar{B} \rangle \Bigr\}\,, \nonumber \\
2m_B r_{13} &= \text{Re}\Bigl\{\langle \bar{B} | \bar{b}_v \,i D_\rho\, (i v \cdot D)\, i   D_\mu \, i   D_\nu \, i   D^\rho \,(-i \sigma^{\mu \nu })\, b_v | \bar{B} \rangle \Bigr\}\,, \nonumber \\
2m_B r_{14} &= \text{Re}\Bigl\{\langle \bar{B} | \bar{b}_v \,i D_\mu \, (i v \cdot D)\, i D_\rho\, i   D^\rho\, i   D_\nu  \,(-i \sigma^{\mu \nu })\, b_v | \bar{B} \rangle \Bigr\} \,, \nonumber\\
2m_B r_{15} &= \text{Re}\Bigl\{\langle \bar{B} | \bar{b}_v \,i D_\mu \, i   D_\nu \, (i v \cdot D)\, i   D_\rho\, i   D^\rho \,(-i \sigma^{\mu \nu })\, b_v | \bar{B} \rangle \Bigr\}\,, \nonumber \\
2m_B r_{16} &= \text{Re}\Bigl\{\langle \bar{B} | \bar{b}_v \,i D_\rho\, i   D_\mu \, (i v \cdot D)\, i   D_\nu \, i   D^\rho \,(-i \sigma^{\mu \nu })\, b_v | \bar{B} \rangle \Bigr\}\,, \nonumber \\
2m_B r_{17} &= \text{Re}\Bigl\{\langle \bar{B} | \bar{b}_v \,i D_\mu \, i   D_\rho\, (i v \cdot D)\, i   D^\rho\, i   D_\nu  \,(-i \sigma^{\mu \nu })\, b_v | \bar{B} \rangle \Bigr\}\,, \nonumber \\ 
2m_B r_{18} &= \text{Re}\Bigl\{\langle \bar{B} | \bar{b}_v \,i D_\rho\, i   D_\mu \, (i v \cdot D)\, i   D^\rho\, i   D_\nu  \,(-i \sigma^{\mu \nu })\, b_v | \bar{B} \rangle \Bigr\} \,.
\end{align}
The LLSA prediction for the parameters~\eqref{eq:ord5pars} is~\cite{Heinonen:2014dxa}
\begin{equation}
    \begin{split}
    r_1 &= \frac{\epsilon_{1/2}^2}{3}(\rho_D^3 + \rho^3_{LS}) + \frac{\epsilon^2_{3/2}}{3}(2\rho_D^3 - \rho^3_{LS}) \,,\\
    r_3 &= -\frac{1}{6}\mu_G^2\rho_{LS}^3 - \frac{1}{3}\mu_\pi^2 \rho_D^3\,,\\
    r_5 &= 0\,,\\
    r_7 &= 0\,,\\
    r_9 &= -\mu_\pi^2\rho_{LS}^3\,,\\
    r_{11} &= \frac{1}{6}(2\mu_\pi^2 - \mu_G^2)\rho^3_{LS} + \frac{1}{3}\mu_G^2 \rho_D^3\,,\\
    r_{13} &= \frac{1}{6}(2\mu_\pi^2 + \mu_G^2)\rho_{LS}^3  - \frac{1}{3}\mu_G^2 \rho_D^3\,,\\
    r_{15} &= 0\,,\\
    r_{17} &= 0\,,
    \end{split}\qquad
    \begin{split}
    r_2 &= -\mu^2_\pi \rho_D^3 \,,\\
    r_4 &= \frac{1}{6}\mu_G^2\rho_{LS}^3 - \frac{1}{3}\mu_\pi^2\rho_D^3\,,\\
    r_6 &= 0\,,\\
    r_8 &= \frac{2}{3}\epsilon_{1/2}^2 (\rho_D^3+\rho_{LS}^3) -\frac{\epsilon_{3/2}^2}{3}(2\rho_D^3 - \rho_{LS}^3)\,,\\
    r_{10} &= \mu_G^2 \rho_D^3\,,\\
    r_{12} &= -\frac{1}{6}(2\mu_\pi^2 + \mu_G^2) \rho_{LS}^3 - \frac{1}{3}\mu_G^2 \rho_D^3\,,\\
    r_{14} &= -\frac{1}{6}(2\mu_\pi^2 -\mu_G^2)\rho_{LS}^3 + \frac{1}{3}\mu_G^2 \rho_D^3\,,\\
    r_{16} &= 0\,,\\
    r_{18} &= 0\,.
    \end{split}
\end{equation}

\bibliographystyle{JHEP} % bst file
\bibliography{refs.bib}

\end{document}